%% file: paper.tex
\documentclass[10pt, sigconf]{acmart}

\usepackage{epsfig,endnotes,times}

\renewcommand\footnotetextcopyrightpermission[1]{} 
\acmConference[arXive]{}{2018}{}
\setcopyright{none}
\usepackage[T1]{fontenc}
\usepackage[utf8]{inputenc}
\usepackage[english]{babel}

\usepackage[smallcaps]{glossaries}
\glsdisablehyper
\usepackage{listings}
\usepackage{graphicx}
\usepackage{lipsum}
\usepackage{subcaption}
\usepackage{capt-of}
\usepackage{enumitem}

\usepackage[binary-units]{siunitx}
\usepackage[vlined,linesnumbered]{algorithm2e}
\SetKwInOut{Input}{Input}
\SetKwInOut{Output}{Output}
\SetKwInOut{Parameter}{Param.}
\SetKwInput{Algorithm}{Algorithm}
\SetKwProg{Fn}{function}{}{}
\SetKwComment{Comment}{{//}}{}

\usepackage{cleveref}
\crefname{section}{§}{§§}
\Crefname{section}{§}{§§}

\usepackage{booktabs}
\usepackage{multirow}
\renewcommand{\arraystretch}{1.2}

\usepackage{afterpage}

\newcommand{\stitle}[1]{\vspace{1ex}\noindent\textbf{#1}}
\newcommand{\inlinecode}[1]{\lstinline!#1!}

\newcommand{\GiB}[1]{\SI{#1}{\gibi\byte}}

\newcommand{\xms}[1]{\SI{#1}{\milli\second}}
\newcommand{\xs}[1]{\SI{#1}{\second}}

\newcommand{\GHz}[1]{\SI{#1}{\giga\hertz}}

\newcommand{\fattree}[0]{Fat-tree\xspace}
\newcommand{\jellyfish}[0]{Jellyfish\xspace}
\newcommand{\jupiter}[0]{Jupiter\xspace}

\newcommand{\timely}[0]{Timely Dataflow\xspace}

\newacronym{SSSP}{SSSP}{single-source shortest-path}
\newacronym{APSP}{APSP}{all-pairs shortest-path}
\newacronym{ONOS}{ONOS}{Open Network Operating System}
\newacronym{SDN}{SDN}{software-defined network}
\newacronym{MPLS}{MPLS}{multiprotocol label switching}
\newacronym{SPF}{SPF}{shortest path first}
\newacronym{SPT}{SPT}{shortest path tree}
\newacronym{OSPF}{OSPF}{Open Shortest Path First}
\newacronym{NFV}{NFV}{Network Function Virtualization}
\newcommand{\pname}{DeltaPath\xspace}

\usepackage{blindtext}

\begin{document}
	
	\date{}
	
	\title[DeltaPath: dataflow-based incremental routing]{DeltaPath: dataflow-based high-performance incremental routing}
	
	\author[D. Dimitrova et. al]{Desislava Dimitrova, John Liagouris, Sebastian Wicki, Moritz Hoffmann, \\ Vasiliki Kalavri, Timothy Roscoe\\
	\small Systems Group, Department of Computer Science, ETH Z\"urich\\ \small firstname.lastname@inf.ethz.ch}
	
	\maketitle
	

	\subsection*{Abstract}
	
	
Routing controllers must react quickly to failures, reconfigurations
and workload or policy changes, to ensure service performance and
cost-efficient network operation.  We propose a general execution
model which views routing as an incremental data-parallel computation
on a graph-based network model plus a continuous stream of network
changes. Our approach supports different routing objectives with only
minor re-configuration of its core algorithm, and easily accomodates
dynamic user-defined routing policies.  Moreover, our prototype
demonstrates excellent performance: on Google’s Jupiter topology it
reacts with a median time of $350ms$ to link failures and serves more
than two million path requests per second each with latency under
$1ms$. This is three orders-of-magnitude faster than the popular ONOS
open-source SDN controller.

	\input{01_intro}
	\input{02_problem}
	\input{03_model}

	\input{04_overview}
	\input{04_implementation}

	\input{04_policy}
	\input{05_setup}

	\input{05_lookups}

	\input{05_failures}
\input{05_switchfailures}
	\input{05_updates}

	\input{05_policy}

	\input{05_bench_parallelism}

	\input{07_deployment}

	\input{02_relatedwork}
	\input{08_conclusion}

	\bibliographystyle{ACM-Reference-Format}
	\bibliography{references}

		
	
\end{document}

%% file: 01_intro.tex
\section{Introduction}\label{sec:intro}

We show how online route computations can be designed and implemented
as incremental streaming computations over graphs.  
We first propose an \textit{abstract execution model} which translates
\gls{SDN} routing logic into a streaming computation. The model is
expressive: it accommodates a variety of routing algorithms with
different cost functions (as used in traffic engineering), or routing
policies (e.g. those used for \gls{NFV} service chaining). It is also
easy to use: it requires only basic declarative configuration to define an
algorithm which is then automatically executed
\emph{incrementally}.  It also leverages the high throughput and low
latency of dataflow frameworks for scalability and performance. 

We then present \pname, a high-performance \emph{incremental
  implementation} of this routing abstraction as a data-parallel
computation in \emph{Timely Dataflow}~\cite{Murray13}, a distributed
execution platform.  Network changes cause \pname to identify only the
affected set of forwarding rules for recomputation.  To our knowledge,
we are the first to propose such an approach for SDN routing
applications.

SDNs offer flexible network control by aggregating network-related
information at a single point. This global view allows for centralized
route computation, replacing traditional distributed routing
protocols. However, scaling centralized route computations to the
size and traffic volumes of datacenters is challenging, prompting 
multi-threaded~\cite{Erickson13,Floodlight} or
distributed~\cite{Berde14,Botelho14,Tootoonchian10} SDN controller design.

While scalability is a challenge, an SDN must also respond quickly to
changes in underlying topology (due to failures or reconfigurations),
traffic policies, and workloads in order to maintain connectivity and
comply with service level agreements: a link failure must initiate
re-routing to avoid traffic loss, or passing a link utilization
threshold should trigger traffic redirection.

Existing SDN routing modules re-execute all required routing
computations from scratch~\cite{Floodlight,ONOS,Opendaylight,Ryu}, but
this is inefficient~\cite{Narvaez00}: even forwarding rules not
affected by the change will be recomputed.  While much research has
looked at algorithms for more flexible traffic control, little
attention has been paid to efficient, incremental execution of those
algorithms in SDN controllers, even though incremental routing
algorithms are well-studied~\cite{iospf,ISIS}.



Here, we show that incremental routing can be combined with the (so
far) separate field of streaming graph computations to provide a
powerful new model for constructing SDN control planes, and point to
a large unexplored design space for network
control systems and their execution models.

\stitle{Contributions.} We make the following contributions.
First, we propose a generic model for executing routing computations
in a pure streaming and incremental fashion (\cref{sec:model}).
Second, we present \pname (\cref{sec:implementation}): a prototype
shortest-path routing application which also supports QoS routing
(\cref{sec:qos_implementation}) and routing policies
(\cref{sec:policy_implementation}).  Third, we conduct evaluations
with realistic-size topologies, including Google's Jupiter
(\cref{sec:evaluation}) and show that \pname computes end-to-end paths efficiently (\cref{sec:eval_lookups}) and recovers extremely
quickly from link and switch failures (\cref{sec:eval_failures}) -- it
is significantly faster (7.9x tail and 30x median latency) than the
popular ONOS controller (\cref{sec:eval_failures}). We also evaluate \pname 's performance for updates in link parameters (\cref{sec:eval_updates}), other routing
strategies, (\cref{sec:qos-eval}), and policies evaluation (\cref{sec:policy_eval}). Finally, we discuss how \pname could be deployed in practice (\cref{sec:deployment}).


%% file: 02_problem.tex
\section{Background and terms}\label{sec:approach}

 
Here we clarify constraint-based routing and other terms we use in the
rest of the paper. 
 
\textit{Constraint-based routing} is a class of routing computations
that comply with (i) administrative policies, and/or (ii)
\emph{Quality of Service} (QoS) requirements \cite{Crawley98,Younis03}
such as bandwidth, packet delay or loss, etc. 

\textit{Policy-based routing}, or policy routing, complies with
high-level policies like those presented in
\cite{fibbing,frenetic,merlin}. We focus on policies which restrict
the type and number of nodes on a path and support service
chaining \cite{Anwer15,Zave17}.  A network administrator who
wants to steer external traffic through a firewall but traffic from
trusted clients should pass untouched, might use a
\textit{waypoint} policy to specify that routes for external traffic
should contain a firewall node, and a \textit{NOT-constraint} policy
to exclude firewalls from paths used by trusted traffic. These
policies are agnostic to the underlying routing algorithm. 

\textit{QoS routing} includes performance targets: video conferencing
requires low-latency paths, while data backup prefers sufficient
bandwidth \cite{Casado14}.  In practice, QoS requirements are met via
different cost functions and path selection criteria.  A link cost
function expresses the cost of a link as a function of a metric
(e.g. bandwidth, load, monetary cost, delay). The costs of all links
comprising a path define a path cost according to an \emph{additive}
or a \emph{min}/\emph{max} path cost function. Finally, a path
selection criterion drives the choice among paths of equal cost. QoS
routing is one tool for traffic engineering, which we discuss in
\cref{sec:related_work}.

\textit{Shortest path routing}, shortest length, or hop-based routing
is the simplest QoS routing algorithm. In shortest path, the cost
is an additive function over constant link costs of 1. Shortest path
is insensitive to traffic dynamics but can be adapted
to express other algorithms. For example, changing the link cost
function to reflect link utilization or residual capacity results in
shortest-distance routing~\cite{Ma97} with bandwidth
constraints. Further replacing the additive path cost function with 
$min()$ results in widest-path routing. We show in
\cref{sec:qos_implementation} how to implement these in our abstract
execution model. Other link attributes can be represented in a similar
way.  We leave enhancements for load balancing~\cite{Gandhi16, He15, Katta16}
and bandwidth provisioning~\cite{Nagaraj16, Tomovic16} for future work.  

%% file: 03_model.tex
\section{Routing model}\label{sec:model}

We now describe our execution model for routing computations: a
graph-based representation of network topology, routing policies,
and flow requests. 

The \textbf{routing model} adopts a dataflow abstraction for routing
computations with three types of inputs represented as event streams:
(i) \emph{topology changes}, (ii) \emph{policies}, and (iii)
\emph{path requests}. The dataflow continuously performs three types
of computation, reacting to the input events it receives.  

First, it uses the stream of topology changes to produce a set of
\emph{base} forwarding rules, according to a QoS routing
strategy (e.g. shortest path, widest path, least-delay path, etc.). Second, it combines the policy stream and the base
rules to produces the \emph{policy} forwarding rules which together with the base rules form the \textit{full} set of forwarding rules. We discuss how his set of rules can be deployed in the dataplane in Section\cref{sec:deployment}. 
Third, for each path request, it generates a routing path by
traversing the full set of rules. Path requests enable network behavior troubleshooting and complement dataplane-driven routing (cf.~\cref{sec:eval_lookups}). The overall dataflow consists of three generic operators as shown in Figure~\ref{fig:dataflow_model} (top).



The \textbf{topology representation} used for routing computations is an undirected\footnote{We assume duplex links for the ease of presentation. Extending the network model to support unidirectional links is straight-forward and requires no changes to \pname's functionality.} property graph $G=(V,E)$ where:
\begin{itemize}
	\item $V$ is a set of nodes, each with a unique \emph{NodeID}, a
          \emph{label} giving the node type (e.g. ``switch'', ``server'', etc.),
          and zero or more \emph{properties} relevant to routing, such
          as the online status of a switch.
	\item $E$ is a set of edges representing physical links. An edge is identified by its endpoints and has associated \emph{properties} (such as link capacity, current utilization, average delay, etc.), 
	a \emph{weight} corresponding to the link's cost (usually computed according to a function), and a $delta$ value $\delta \in \{-1,+1\}$, used to represent changes in $G$ (explained below).  
\end{itemize}


As the network topology changes, $G$ evolves correspondingly.
Here, we consider three types of change: (i) \emph{adding or removing links},
(ii) \emph{adding or removing nodes}, and (iii) \emph{updating link weights}.  
Changes in node and edge properties can be handled similarly.

Link additions and removals are represented by adding edges with delta
values equal to +1 ($\delta = +1$) and -1 ($\delta = -1$),
respectively. Such edge collections are added to $G$ and 
result in the removal of existing edges by aggregating the deltas
of edges with common attributes and discarding all edges whose
deltas sum to zero. For instance, a link between switches $s_1$ and
$s_2$, represented by ($s_1$, $s_2$, $weight$, $\delta=+1$), is
removed by adding an edge ($s_1$, $s_2$, $weight$,
$\delta=-1$). 
Adding or removing a node $n$ (e.g. a switch) is reduced into adding or removing all edges with $n$ as an endpoint.  
Finally, updates to link weights are modeled by removing the corresponding edge, followed by adding a new edge with the same attributes and the updated weight.
This delta-based approach is reminiscent to the one used in \cite{loo2006} and other view maintainance approaches in databases, and enables asynchronous incremental computations, as we show in \cref{sec:implementation}. 

\textbf{Policies} express constraints on the nodes appearing in a path.  
Adopting the notation in \cite{frenetic}, a policy $pol::=S:c_i(n):T$ is one or more \textit{node} constraints $c_i(n)$ on a path between an origin node $S$ and a target node $T$. Node IDs and IP address are two possible ways to identify $S$, $T$ and $n$. In the example of Section~\cref{sec:approach}, the waypoint policy is written as $pol\ ::=\ S_{ext} : n_{FW}: T_{int}$, where $S_{ext}$ is an external node, $T_{int}$ is an internal node and $n_{FW}$ is a firewall node. A NOT-constraint denoting that a flow must not go through a firewall is expressed as $pol\ ::=\ S_{trust} :\ !n_{FW} : T_{int}$, where $S_{trust}$ is a node registered with a trusted client. Policies are added or removed via the policy operator of Figure~\ref{fig:dataflow_model}. 

The model also supports path policies such as the ones proposed in~\cite{fibbing}, e.g., single failover path $pol\ ::=\ S: b_{path}:T$, 2-way multipath $pol\ ::=\ S: p_1 \lor p_2:T$, and redundant transmission over two paths $pol\ ::=\ S: p_1 \land p_2:T$. The model complements switch-based access policies by providing forwarding rules for whitelisted traffic\footnote{If blacklisting is used this can be handled directly at the switch.}.
Other types of policies, such as those with general path constraints
in the form of regular expressions and complex multipath policies, are beyond the scope of this paper and we leave them as future work. 

\textbf{Path requests} represent either flow requests or troubleshooting queries. The former exhibits if not all flows can be handled at the switch due to limited memory. The latter lets us retrieve the path a flow should have taken and use it to troubleshoot its exhibited routing behavior.  
A request has the form $flow::= (flowID, S, T)$, where $flowID$ is an ID that uniquely identifies a flow, $S$ and $T$ are the flow's origin and target nodes respectively.

%% file: 04_overview.tex
\section{\pname: dataflow-based routing}\label{sec:implementation}

We now present \pname's design and detail the functionality and
implementation of its core components. 


\begin{figure}
	\centering
	\includegraphics[width=.85\columnwidth]{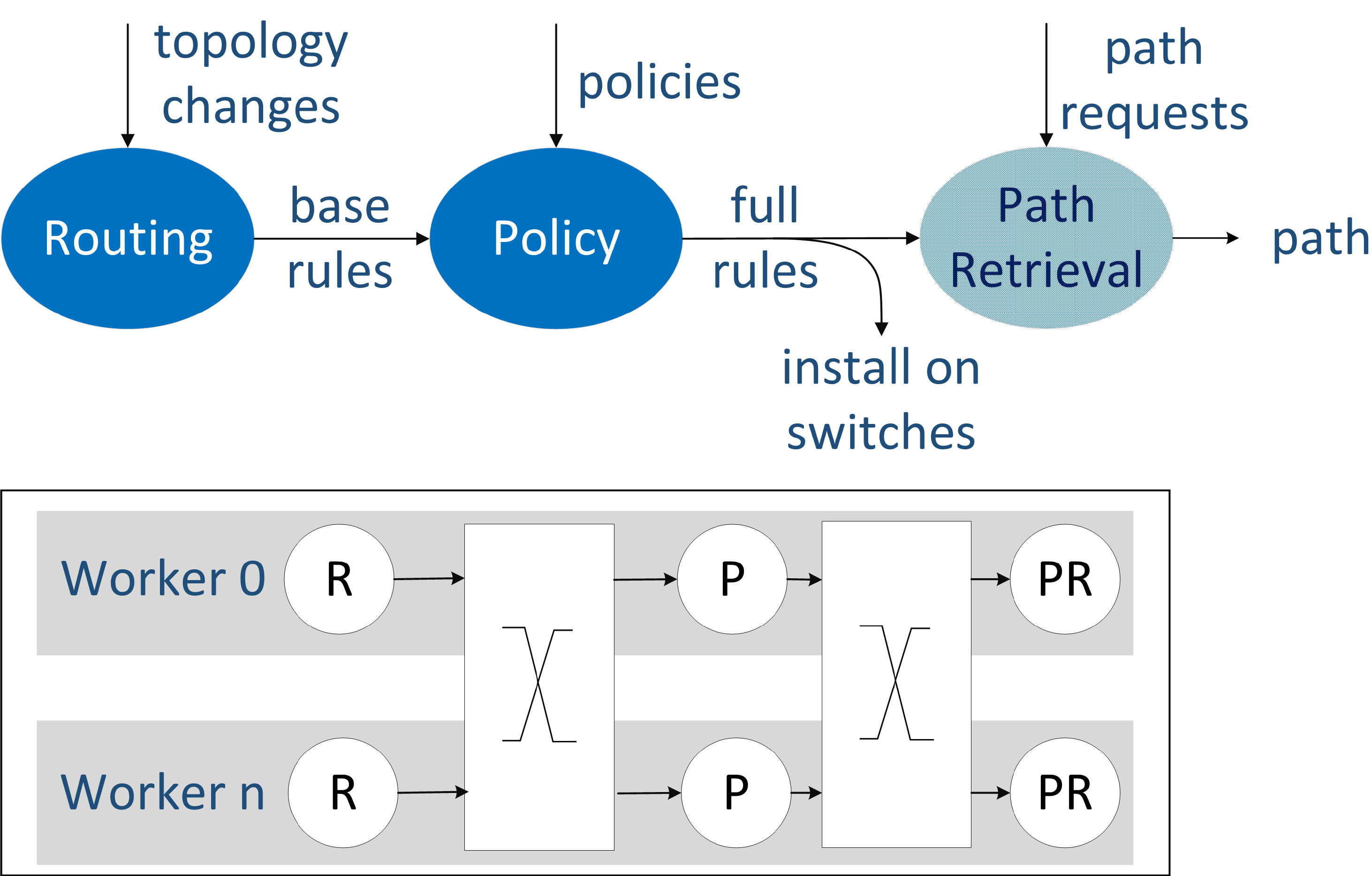}
	\caption{\small \sl \pname's logical dataflow with three operators (top) and physical dataflow (bottom). Rectangles in the physical dataflow denote data shuffle stages.}
\label{fig:dataflow_model}
\end{figure}


\pname leverages the global network view of SDNs to first
\emph{proactively} compute forwarding rules for all pairs of network
switches according to a QoS routing strategy. This happens in the routing operator of Figure~\ref{fig:dataflow_model} during \pname's
initialization on an initial snapshot of the graph $G$
(cf. \cref{sec:model}). To minimize disruption of switch operations forwarding rules (and their policy-derived counterparts) are pre-installed at switches. \pname thereafter employs efficient incremental computations to \emph{reactively} update rules on need-to basis, e.g., upon a network or a policy change.
\pname's approach applies to wide range of routing strategies (cf. \cref{sec:qos_implementation}) and is agnostic to the granularity of routing decisions, e.g., per flow, flow class or ToR switch (cf. \cref{sec:deployment}).

Each operator in \pname is implemented as a streaming operator above \timely,
first introduced in Naiad~\cite{Murray13}.  \timely is an ideal
platform for us due to (i) its \textit{event-driven} programming
model, which allows us to naturally represent streams of asynchronous
routing requests and network updates, and (ii) its native support for
arbitrary \emph{iterative computations}, common to routing
tasks. Furthermore, in our experience, Timely's Rust implementation
outperforms other platforms such as Flink \cite{flink} and Spark
Streaming \cite{spark}.


\timely employs a data-parallel execution model and processes the dataflow graph
across a configurable set of \textit{workers} as shown in the lower part of 
Figure~\ref{fig:dataflow_model}. Each worker processes a partition of the input streams (\emph{data parallelism}) asynchronously and exchanges data with other workers via channels. All records are tagged with logical timestamps (\textit{epochs}) to keep track of
workers' progress and ensure the consistency of the asynchronous
computation. 
asynchronous: each worker periodically issues statements about local
work, together comprising a global view of progress.


\subsection{QoS routing in \pname}\label{sec:qos_implementation}

\begin{figure}[t]
	\centering
	\includegraphics[scale=0.25]{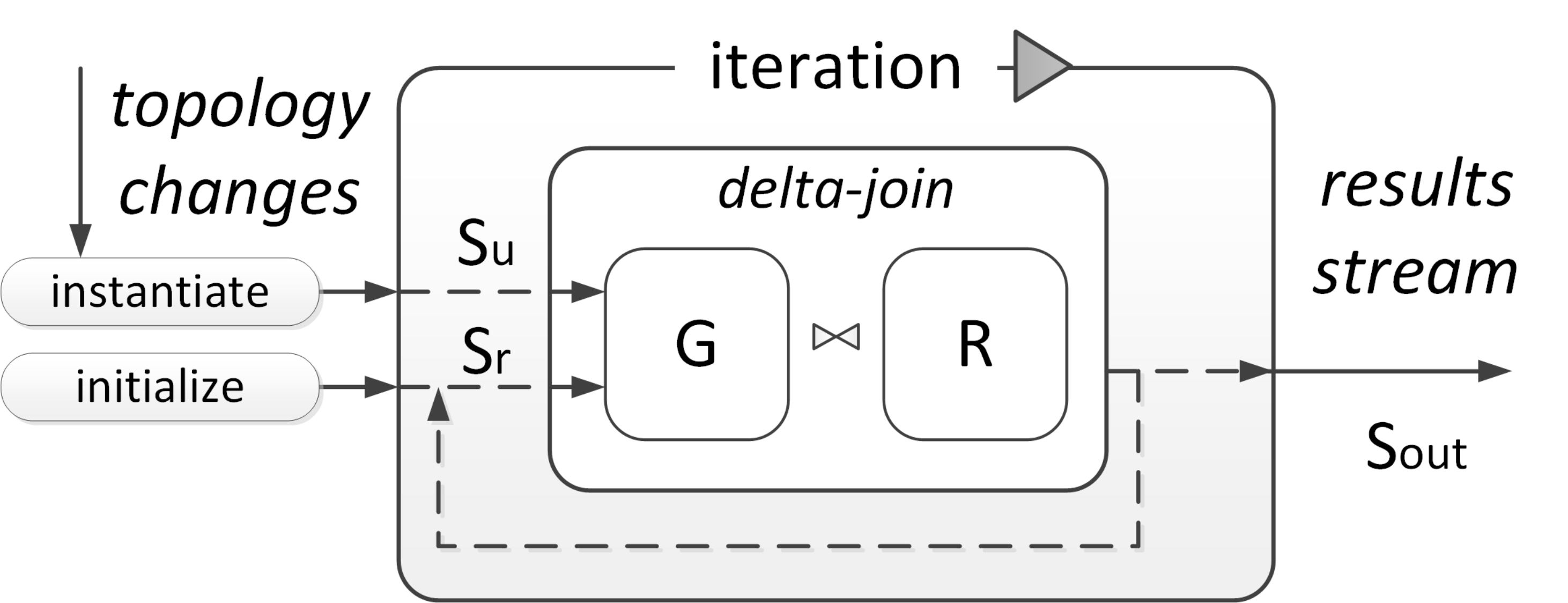}
	\caption{\small \sl \pname routing as a cyclic dataflow computation.}
	\label{fig:apsp-operator}
\end{figure}

The routing operator (Figure~\ref{fig:apsp-operator})
implements a QoS routing strategy as a continuous cyclic computation accepting two 
inputs: a stream of topology changes ($S_u$), represented as edges (cf. \cref{sec:model}), and a
stream of forwarding rules ($S_{r}$), connected in a feedback loop with the operator's output. 
The output is a stream $S_{out}$ of base forwarding rules.

A network adminstrator deploys a routing strategy by defining three
functions, which are used to instantiate the routing operator:  
a link cost function $f_\ell$, a path cost function $f_p$, and a path selection function $f_s$. 
\pname then transparently provides incrementality.

The routing operator maintains some state: a snapshot of the current network graph $G$ (cf. \cref{sec:model}) and a corresponding collection of base forwarding rules, $R$. 
Specifically, $G$ is a collection of edge records of the form $(src, dst, w, \delta)$, where $src$ and $dst$ are endpoint nodes, and
$w$ is the edge weight given by $f_{\ell}$. Edge records are constructed on-the-fly from incoming network changes by the \emph{instantiate} component of Figure~\ref{fig:apsp-operator}. 
$R$ is a collection of forwarding rules of the form $(src, dst, next, p_{cost}, p_{length}, \delta)$, where $next$ is the next-hop node in the path
from $src$ to $dst$, $p_{cost}$ is the total cost of the
path given by $f_p$, and $p_{length}$ is the length of the path in number of hops. $\delta \in \{-1,+1\}$ is used to express updates to edges and forwarding rules in $G$ and $R$ respectively.
At startup \footnote{A one time event following the deployment of the routing application in a live controller.}, $R$ is populated with initial forwarding rules, called \emph{tautologies}, 
generated by the \emph{initialize} component of Figure~\ref{fig:apsp-operator}.
Table~\ref{tbl:notation} summarizes this notation.

 
\begin{table}[]
	\small
	\begin{center}
	\setlength\tabcolsep{4pt}
	\begin{tabular}{llll}
		$src$      &: source node                  & $S_u$      &: stream of link updates \\ 
		$dst$      &: destination node             & $S_r$      &: stream of rule updates \\ 
		$next$     &: next-hop node                & $G$          &: set of links (edges)  \\ 
		$w$        &: link weight (cost)                  & $\Delta G$ &: set of link changes    \\ 
		$p_{cost}$  &: path cost                   & $R$          &: set of rules   \\ 
		$p_{length}$  & : path length (hops) 	   & $\Delta R$	   &: set of rule changes   \\ 
		$f_p$   &: path cost function &  $f_s$   &: path selection function   \\ 
		$f_\ell$   &: link cost function &  $\delta$ &: delta value ($\delta\in\{\-1,+1\}$)  \\ 
	\end{tabular}
	\end{center}
	\caption{\small \sl Notation of Section \ref{sec:implementation}}
	\label{tbl:notation}
\end{table}

\stitle{Cost functions.}
The link cost function $f_\ell$ defines edge weights, which can be either static, as defined by hop count, link capacity or propagation delay, or dynamic (updated by network monitoring), as defined by link utilization or observed delay. Since the choice of $f_\ell$ does not affect the routing computation, \pname easily extends towards multi-parameter routing strategies, e.g, based on capacity, delay and buffer space \cite{GuJ16,WiR94}.

The path cost function $f_p$ defines the cost of a newly discovered path that is constructed at a specific point during the iteration by `expanding' a previously computed path with an adjacent edge.
$f_p$ is applied on the cost of the existing path ($p_{cost}$) and the weight of the edge ($w$). 
The function $f_s$ selects the best path (among all discovered) for each $(src,dst)$ pair. This is an aggregation function applied to the collection $R$ of forwarding rules in the local state of the routing operator.
While $f_p$ can be an additive or a min/max function, $f_s$ is typically a min/max.

Table~\ref{tbl:functions} shows the functions for shortest path,
widest path, and two versions of the shortest-distance algorithm  
(one with weights denoting links' available bandwidth, and another
with weights denoting links' utilization). 

Note that in real networks, different classes of traffic coexist, requiring different QoS routing strategies. In the previous
example, two reasonable strategies could be a shortest-distance
strategy with delay-based weights for video conference flows, and a
widest-shortest path srtategy for backup application flows. This could be realized using two instances of \pname, each one running its own strategy in parallel. This approach provides a simple solution to traffic differentiation and could be optimized by leveraging knowledge on traffic patterns.

\begin{table*}[t]
	\scriptsize
	\centering
	\setlength\tabcolsep{2pt}
	\renewcommand{\arraystretch}{1.5}
	\begin{tabular}{|c|c|c|c|}
		\hline
		\textbf{QoS routing algorithm} & \textbf{link cost function} & \textbf{path cost function} & \textbf{path selection function*}  \\ \hline\hline
		hop-based			& $w$ = 1 & $f_p$ = $1+p_{cost}$  & \multirow{2}{*}{ \multirow{4}{*}{$f_s$ = \texttt{min($p_{cost}$)}}}  \\ 
		\cline{1-3}
		shortest-distance   & \multirow{2}{*}{$w$ = $\dfrac{1}{\textnormal{link's free bandwidth}}$} & \multirow{4}{*}{$f_p$ = $w+p_{cost}$}   &    \\ \
		(free bandwidth) & & & \\
		\cline{1-2}
		shortest-distance & \multirow{2}{*}{$w$ = $\textnormal{link's utilization}$} &  &     \\ 
		(link utilization) & & &\\
		\hline
		shortest-widest & $w$ = $\textnormal{link's free bandwidth}$  & $f_p$ = \texttt{min\{$w$,$p_{cost}$\}} & $f_s$ = \{\texttt{max($p_{cost}$),min($p_{length}$)}\}    \\ 
		\hline
		\multicolumn{4}{l}{* The path selection function $f_s$ is applied to all rules for a $(src,dst)$ pair in the collection $R$ (cf. Figure~\ref{fig:apsp-operator})}
	\end{tabular}
	\caption{\small \sl Example cost functions for QoS routing in \pname}
	\label{tbl:functions}
\end{table*}

%% file: 04_implementation.tex
\subsection{Incremental routing computation} \label{sec:net_algorithm}

For simplicity we describe \pname's implementation assuming a
shortest-distance routing strategy, where a path $p$ between two
nodes $src$ and $dst$ in the network graph minimizes the
sum of weights (e.g. utilization) on its edges.  
Other routing strategies are implemented by providing the appropriate
cost functions of \cref{sec:qos_implementation}. 

The routing operator is, in database terms, a \textit{delta-based
  join} of $G$ and $R$ at a specific point in time (an \emph{epoch} in
Timely Dataflow).  
It implements a data-parallel incremental \emph{label propagation} algorithm
as a streaming dataflow operator.  
Label propagation is an iterative procedure whereby at each step every
node receives values (labels) from its neighbors, applies an update
function to its own label, and propagates this result back to its neighbors.

\stitle{The algorithm} works in two steps. First, an initial label propagation 
over the topology yields a complete set of forwarding
rules (the base rules) according to a routing strategy, and thereafter the same algorithm 
incrementally reacts to changes in the topology and maintains these rules.

Algorithm \ref{alg:apsp} shows the routing logic applied on a per-epoch basis. 
Assume that an initial network topology exists in $G$, 
  and the set of tautology rules is generated and pushed to $S_r$.
  For a shortest-distance strategy,
the tautology rules have the form $\{n, n, n, 0, 0, 1\}$ for each
network node $n$ -- intuitively, ``each node can reach itself via
itself at zero cost''.

Lines \textbf{3-8} process incoming forwarding rules in
$S_r$. The algorithm first collects the updates to $R$ (line \textbf{4}) in a collection $\Delta R$ (initially all tautologies)
and joins $\Delta R$ with $G$ as follows: For a rule $r \in S_r$ and an edge $l
\in G$, the join is performed on the ID of the source nodes, $r.src=\ell.src$, and produces new forwarding rules $r'$ of the form $(l.dst,
r.dst,$ $r.src, l.w, l.w + r.p_{cost}, l.\delta * r.\delta)$.\footnote{ Semantics here imply that the edges in $G$ are \emph{symmetric} (cf. \cref{sec:model})}
Intuitively, a new rule $r'$ indicates 
that the destination node of the link $\ell \in G$ can reach the
destination node of the rule $r \in \Delta R$ through $r.src$ (which
is the same as $\ell.src$, hence, adjacent to $\ell.dst$) and with a
cost defined by the path cost function $f_p(\ell.w,r.p_{cost}) = \ell.w + r.p_{cost}$
($r.p_{cost}$ is the cost of rule $r$ that amounts to the aggregated edge weights in the current path between $r.src$ and $r.dst$).

Lines \textbf{3-8} execute repeatedly. After each iteration, the
new forwarding rules are pushed
to $S_r$ (line \textbf{8}). These are used at the start of the next
step to update the $R$ (line \textbf{4}) using the path selection
function $f_s(R) = min(p_{cost})$, which selects the best forwarding
rule (path) for each distinct ($src, dst$) pair in $R$.  

\begin{algorithm}[t]
\small
\Algorithm{\textsc{\pname routing}}
\Input{A stream $S_{u}$ of network updates and a stream $S_{r}$ of forwarding rules}
\Output{A stream $S_{out}$ of forwarding rules}
\BlankLine
\Comment{The state (initialized once)}
        
let $G$ be the network graph 
        
let $R$ be the collection of forwarding rules
\BlankLine

\Comment{Process stream of forwarding rules}
\While{there are forwarding rules in the input stream $S_{r}$} {
                    
                    
                    
                    Update local state $R$ and collect changes, let $\Delta R$;
                    
                
                    \For {each pair $(r,\ell)\ |\ r \in \Delta R,\ \ell \in G : r.src = \ell.src$} {
        
                    	Create a new forwarding rule $r'$ as:
	
			$r' = \{\ell.dst, r.dst, r.src, \ell.w + r.p_{cost}, \ell.\delta * r.\delta\}$;
	
			Push $r'$ to the rule stream $S_{r}$;
                    }            
  }
  
 \Comment{Process updates coming from the network}  
  \While{there are edge records in the input stream $S_{u}$} {
                    
                    
                    Update local state $G$ and collect changes, let $\Delta G$;
                    
                    \For {each pair $(r,\ell)\ |\ r \in R,\ \ell \in \Delta G : r.src = \ell.src$} {
        
                    	Create a new forwarding rule $r'$ as:
	
			$r' = \{\ell.dst, r.dst, r.src, \ell.w + r.p_{cost}, \ell.\delta * r.\delta\}$;
	
			Push $r'$ to the output stream $S_{r}$;
                    }            
  }
%
%
%
%
%
%

  \caption{Core routing logic in \pname}\label{alg:apsp}
  \vspace{0.3cm}
\end{algorithm}



Lines \textbf{9-14} describe the processing of incoming network changes in $S_u$.
The algorithm first collects updates to $G$ in a collection $\Delta G$
(line \textbf{10}). During this step, it matches $S_u$ records against
edges in $G$ and sums $\delta$ values of edges with the same $(src, dst, w)$.
All records with $\delta=0$ are then
garbage-collected. This process guarantees a consistent view of the
topology graph where the network updates are reflected on the topology
before the incremental rule computation begins. 
Then, the algorithm joins $\Delta G$ with $R$ on the ID of the source nodes, $r.src = \ell.src$, to generate new forwarding rules. 
The latter are pushed to $S_r$ and drive the incremental computation in lines \textbf{3-8}.
The algorithm uses the $\delta$ values of matched
records to identify rules to recompute.  Specifically, it multiplies
the $\delta$ values of matched records (lines \textbf{7} and \textbf{13}) so that resulting records with
$\delta= -1$ cancel out existing invalid rules, while rules with $delta
= 1$ and the next minimum $p\_cost$ become the new established
rules. At each step, rules with $\delta = 0$ are garbage-collected as with edges.

The computation converges when $\Delta R$ and $\Delta G$ are both empty, so the current epoch has been processed. 
At that point, the updated forwarding rules are pushed to the output stream $S_{out}$. 
These rules correspond only to the updates applied to $R$ with respect to the previous epoch.

%

\begin{figure}[t]
	\centering
	\includegraphics[scale=0.3]{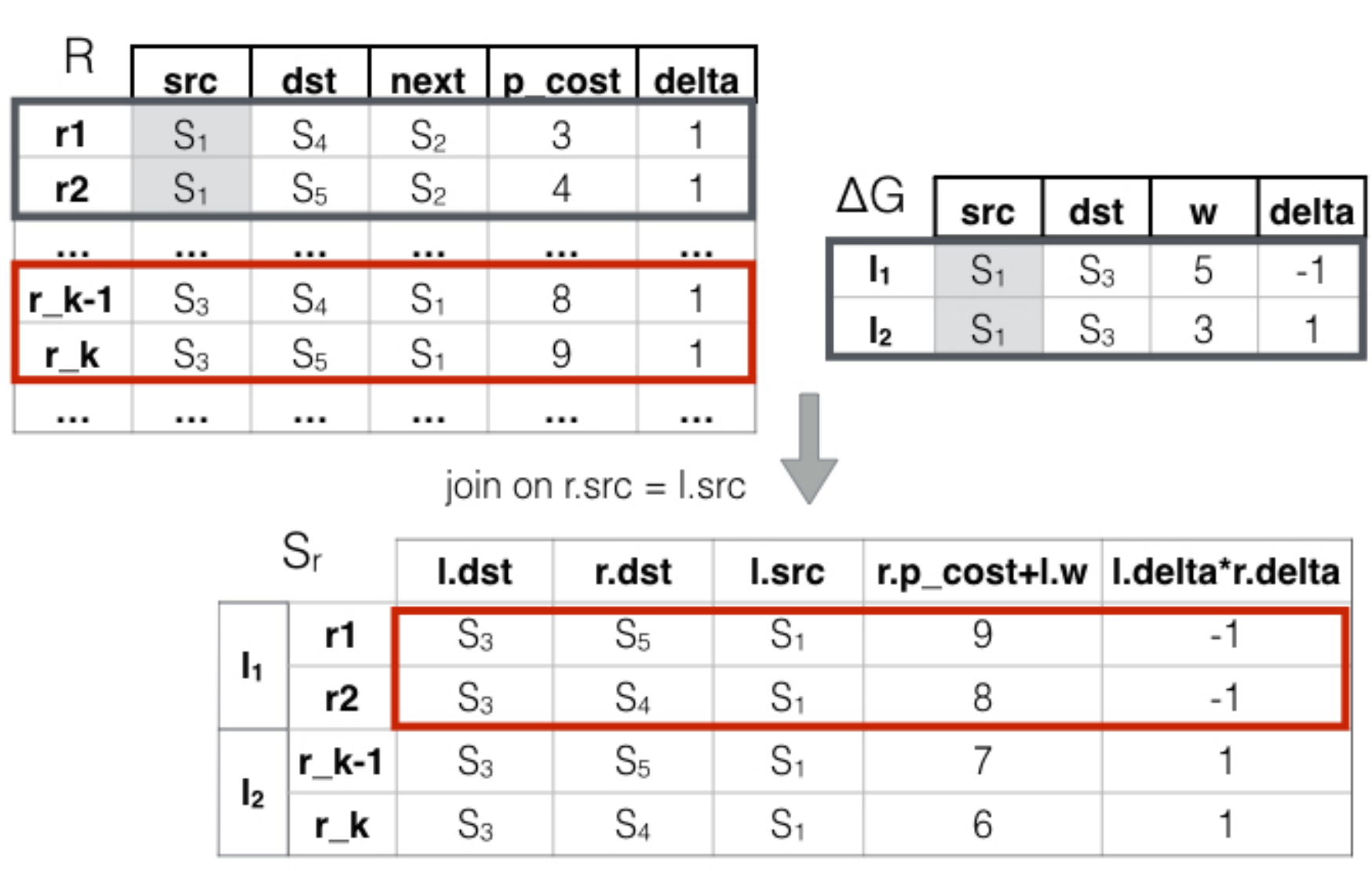}
	\caption{\small \sl An example of incrementally updating forwarding rules. The top two rules in the result table match the existing invalid rules $r_{k-1}$, $r_{k}$. When updating $R$, the routing algorithm will sum-up their $\delta$ values and garbage-collect them.
	}
	\label{fig:incremental-rule-example}
\end{figure}

\stitle{Example:} 
Table $R$ shows a hypothetical collection of forwarding rules after the shortest-distance algorithm has
converged ($p_{length}$ attribute omitted). For simplicity, we only show two rules for nodes (switches) $S_1$ and
$S_3$. Table $\Delta G$ contains two 
example updates: a removal of link $(S_1, S_3, 5)$ and a 
new link between $S_1$ and $S_3$ with weight 3. In this case, 
Algorithm~\ref{alg:apsp} (lines \textbf{9-14}) will  join
 $R$ and $\Delta G$ to produce the records in the bottom table
of Figure~\ref{fig:incremental-rule-example}, which are pushed to the rule stream $S_r$. Note that the top two
rules in this table match the existing, now invalid, rules $r_{k-1}$
and $r_{k}$ in $R$. When the latter is updated (line \textbf{4}), the algorithm will sum their
$\delta$ values and garbage-collect them.

\stitle{Discussion:}
\pname's routing component \textit{does not materialize
  any path}; instead, it computes per-hop forwarding rules for the
paths between all pairs of reachable nodes. 
Path construction is done lazily by the policy and the path retrieval components, as we explain in \cref{sec:path_retrieval} and \cref{sec:policy_implementation}. 

Algorithm~\ref{alg:apsp} can be regarded as a specialization of Differential Dataflow \cite{differential}, 
a library built on top of Timely Dataflow to automatically incrementalize computations in the spirit of \pname. 
To further explore the potentials of our approach, we have also implemented a prototype on Differential Dataflow, however,  
performance in that case degraded by $8\times$ due to the different (internal) data structures used by \cite{differential-rust}.

As a final note, \pname's routing component could be made fault-tolerant via standard
replication approaches but since \timely already distributes transparently, more
sophisticated fault-tolerance techniques~\cite{jacques2016,carbone2015} are applicable. We leave this as future work.

%% file: 04_policy.tex

\subsection{Policy evaluation} \label{sec:policy_implementation}

\pname's policy component allows network operators to define \emph{custom} routing policies. 
Recall that we only consider two types of policies here: (i) waypointing policies of the form $pol\ ::=\ S : c_i(n): T$, and (ii) NOT-constraint policies of the form $pol\ ::=\ S : !n : T$ (cf. \cref{sec:model}). 
 
\stitle{Waypointing policies.} 
Waypointing policies are evaluated by the policy component in two steps. First, the input policy is parsed to create a tree structure of constraints. 
Then, the component traverses the tree and initiates a series of path retrievals conforming with the policy. 
The example policy $pol\ ::=\ S_{ext} : n_{FW}: T_{int}$ from \cref{sec:model} breaks down into two path retrievals: one from $S_{ext}$ to $n_{FW}$, and a second one from $n_{FW}$ to $T_{int}$. 
In general, for a waypointing policy with $k$ constraints $(c_1,c_2,...,c_k)$, the policy component triggers $k+1$ path retrievals. The procedure is identical to the one described in \cref{sec:path_retrieval} but executes on the set of base forwarding rules. The resulting policy rules are combined with the base rules and pushed to the path retrieval component. 
Intuitively, a collection of rules for a waypointing policy is a subset of the base rules generated by the routing component for the current graph snapshot.

\stitle{NOT and path-constraint policies.} 
\pname's general model (cf. \cref{sec:model}) provides a very simple approach to implementing NOT and path-constraint policies.
A NOT-constraint can be regarded as a node failure, while a path-constraint can be simulated by a link failure.
As an exmaple, consider the policy $pol\ ::=\ S_{trust} : !n_{FW} : T_{int}$. 
First, a request to remove $n_{FW}$ from $G$ is sent to the routing component. This removal takes place in a copy of the current graph snapshot so that other active policies are not affected.
Then, the updated base rules are sent to the policy component and a policy-compliant path between $S_{trust}$ and $T_{int}$ is retrieved and pushed to the output. 
The copy of $G$ associated with the specific NOT-constraint policy is kept separately in the state of the routing operator, and is updated by the incoming network changes independently from the original graph $G$. Policies with path constraints are supported in a similar fashion. For example, to find a backup path \pname (1) looks up the links consisting the main path and (2) removes those links from a copy of the topology graph. The recomputed forwarding rules deliver a link disjoint backup path. The backup path enables policies on 2-way multipath and redundant transmissions as well.

Despite its simplicity, this approach might not scale well for very large networks. One direction we are currently exploring to further optimize implementation is reusing common forwarding rules across the different copies of $G$.


\subsection{Path retrieval} \label{sec:path_retrieval}

Given the full set of forwarding rules, a path retrieval is reduced into a simple \emph{pointer chasing} using the $next$ attribute of each rule. 
For a source node $s$ and a destination node $d$, the path retrieval starts with a lookup in the collection of rules to retrieve a rule, let $r_1$, whose $src = s$ and $dst = d$. Then, it repeats the lookup in the rule collection with $src=r_1.next$ and $dst=d$, retrieving a second rule $r_2$. The procedure iterates for $k$ steps until it finds a rule $r_k$ such that $r_k.next=d$. The set of forwarding rules $r_1...r_k$ constitute the shortest path between the source $s$ and destination $d$. 

The aforementioned pointer chasing is essentially the functionality of \pname's path retrieval component, applied to its collection of policy rules upon a flow request. Path retrievals are also performed during the evaluation of policies as we explain in the next section.


%% file: 05_setup.tex
\section{Evaluations}\label{sec:evaluation}

We evaluate \pname's performance and compare it with the
state-of-the-art and widely-deployed ONOS SDN controller -- a choice
we discuss below. We are particularly interested in the following metrics:
\begin{itemize}[leftmargin=*]
	\setlength\itemsep{0em}
	 \item policy evaluation: \pname computes a waypointing policy with 5 nodes in 3ms and evaluates NOT-constraints in comparable time to node failures.
	\item recovery latency from \textit{isolated} link and switch
          failures: we show that \pname recomputes routing state
          \textit{30 times faster} than the ONOS controller.
	\item recovery from successive failures: \pname's on-demand
          recomputation is stable under a sequence of failures,
          whereas ONOS' recovery time quickly degrades due to
          expensive precomputation operations.
\end{itemize}  

We focus  on link removals since algorithmically they are
equivalent to additions, but represent failures and other unscheduled
events of high importance in production networks.


We run \pname and ONOS on a single quad-socket Intel Xeon
E5-4640 running Debian Jessie (8.2) with \GiB{512} RAM, and
frequency scaling enabled, and 8 \GHz{2.40} cores (16 threads) per socket.  Our
compiler is \texttt{rustc} 1.20.0-stable. 


\stitle{Topologies.} We evaluate against the four topologies in
Table~\ref{tab:topology-params}.  
\fattree \cite{FatTree} is a version of the common leaf-spine
structure which maintains multiple paths between any two access
switches and consists of multiple \textit{pods} -- sets of access and
aggregation switches.  \jellyfish \cite{Jellyfish} is a low-diameter
random graph-inspired network designed for easy extendibility, and
sized here to support the same number of hosts as \fattree. 
Topo-R is an older topology from a real, operational
industrial datacenter which has evolved over time and so has irregular
structure compared to a tree topology. 
\jupiter is Google's scalable datacenter
fabric~\cite{Jupiter}, whose building blocks consist of
Top-of-Rack, aggregation, and spine layers. Blocks at the upper two layers
resemble 2-stage blocking network. 

\begin{table}
	\small
	\begin{tabular}{@{}cccccc@{}} \toprule
		Topology & {Hosts} & {Switches} & {Ports} & {Links} \tabularnewline
		\midrule
		\jellyfish & 27648 & 1280 & 48 & 6912 \tabularnewline
		\fattree & 27648 & 2880 & 48 & 55396 \tabularnewline
		Topo-R  & 19404 & 546 & * & 917 \tabularnewline
		Jupiter & 98304 & 5632 & 64 & 87040 \tabularnewline
		\bottomrule
	\end{tabular}

	{\footnotesize * The number of ports per switch in Topo-R varies.}
	\caption{\small \sl Network topologies used in the experiments}
	\label{tab:topology-params}
	\vspace{-0.5cm}
\end{table}

\stitle{Link cost.} We evaluate with two link weight assignments. The
\texttt{Hop-count} plan gives all links an equal weight of 1, as is
common in datacenter networks, and effectively represents hop-based routing. The \texttt{Uniform} plan corresponds to shortest distance routing based on  weights represent link's utilization, distributed in the range [1,100] with 100 denoting the full link bandwidth. To simulate utilization changes we introduce a \textit{random weight update} as an arbitrary increase or decrease in utilization for a randomly chosen link. We also introduce \textit{batch size} as the number of updates processed concurrently to reflect dynamic datacenter workloads. 
We note that shortest distance algorithms are more efficient in link utilization \cite{Ma97} than their widest-path counterparts.
This choice of algorithms allows us to demonstrate both the applicability of our approach and its generality for constraint routing. 


\stitle{Comparing to existing controllers.}
Comparing \pname to existing SDN controllers is a challenge, since
they are not designed for performance measurement or analysis, and
use widely different interfaces to express routing policy.
We considered the \gls{ONOS} \cite{ONOS}, OpenDaylight~\cite{Opendaylight},
Ryu~\cite{Ryu}, and OpenMUL~\cite{OpenMUL}.  With the exception of
OpenMUL, which uses \gls{APSP}, all perform \gls{SSSP} (most often
Dijkstra's algorithm) on a per-flow basis.  We are unaware of any 
open-source SDN controllers that perform \textit{incremental}
\gls{APSP} computation.  We chose \gls{ONOS} based
on performance and market dominance~\cite{ODL_report, ONS_report}.


\gls{ONOS} expects flow demands to be specified declaratively
via \emph{intents}.  For example, a \texttt{PointToPointIntent} specifies that two hosts engage in communication. Intents can be \emph{unprotected}, served by a single shortest path between the endpoints,
or \emph{protected}, having a second, \emph{node-disjoint} path as backup. 
When a link failure occurs, all affected intents are looked up in an
index.  Unprotected paths are recalculated by running Dijkstra on the
updated topology graph. Protected paths fail over quickly via stored backups
but then
require running Suurballe's algorithm \cite{Suurballe} again to restore redundancy.  

As we show, ONOS is slow in handling link updates because those are
interpreted as network failures and trigger full path recomputation on a
per-intent basis.


%% file: 05_lookups.tex
\subsection{Retrieving end-to-end paths}\label{sec:eval_lookups}

Before presenting the core evaluations we discuss \pname's performance in computing a \emph{path}. Path retrieval is essential in supporting policies (cf. \cref{sec:policy_eval}) and \pname's efficiency lays in extracting the path from a precomputed set of individual forwarding rules\footnote{This is in contrast to less efficient approaches that either precompute and materialize routing paths or trigger path computations on the fly.}.  

\pname can return a single path in $0.01ms$, \emph{five orders of
  magnitude faster than ONOS}, and can answer \emph{2.3 million path
  requests per second} on Jupiter.


After initializing a network with the set of base forwarding rules,
we generate batches of randomly selected $(src,dst)$ path requests and
increase the batch size $k$ from 1 to 8192 in powers of
2).  For each $k$ we submit 500 batches to a single \pname
worker\footnote{This is a scenario where parallelism is undesirable
  due to communication overhead, since path lookup is cheap in
  \pname.}. We measure how long it takes to return all the paths in
each batch.  No changes to topology or link weights occur in these
experiments.

\begin{figure}
	\centering
	\includegraphics[width=.9\linewidth]{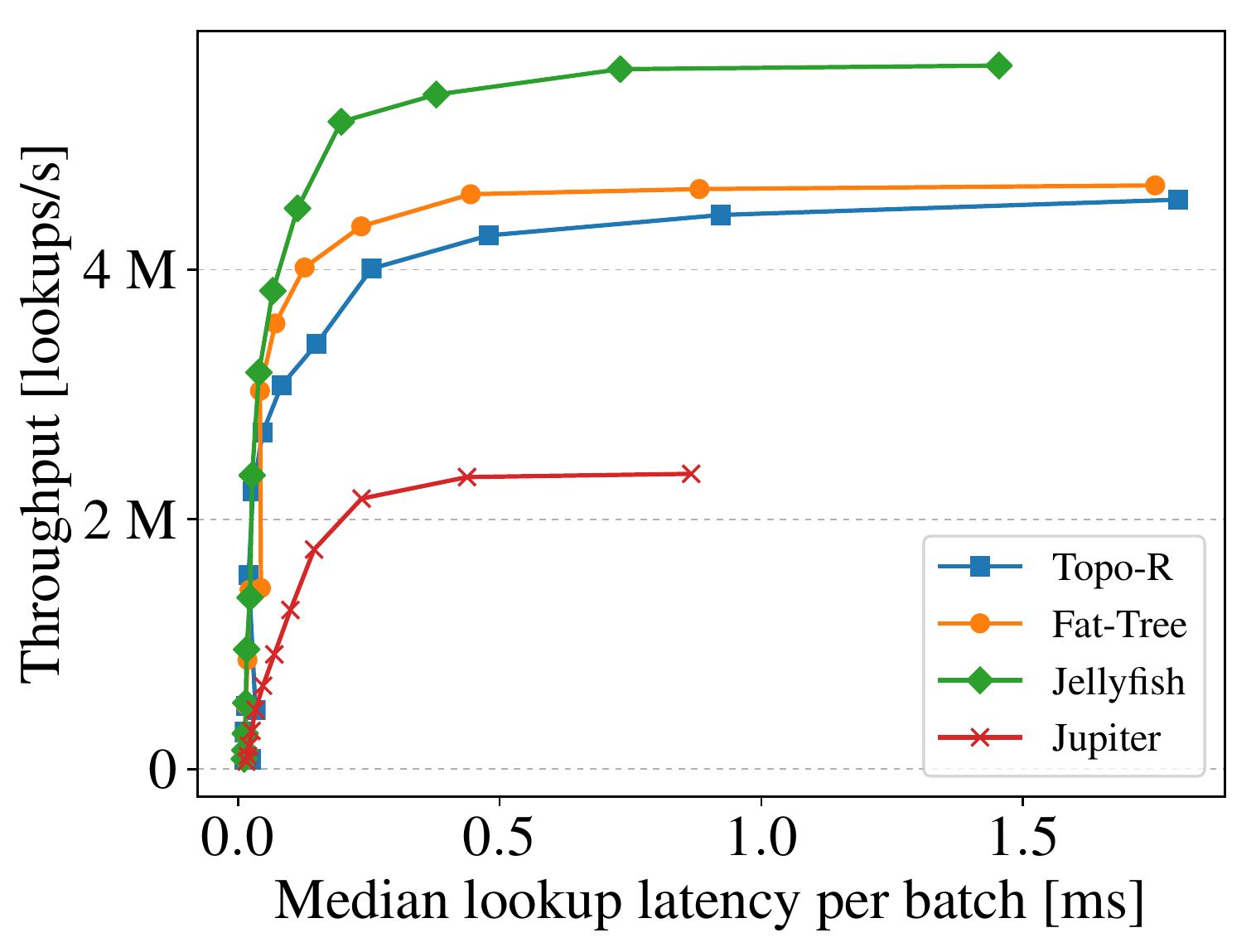} 
	\captionof{figure}{\small \sl Throughput vs Latency for path lookup, \texttt{Hop-count} plan}
	\label{fig:lookups-hopcount}  
\end{figure}


Figure~\ref{fig:lookups-hopcount} shows
throughput vs. latency for the different batch sizes in the \texttt{Hop-count plan}.  
\pname can comfortably retrieve more than 1M paths for all topologies
in less than $1s$. Using uniformly random distribution of link weights (\texttt{Uniform plan}) results in similar behavior at higher latency and we omit it due to space limits. 

Differences in the average path length explain
the relative performance for lookups for the different
topologies. Intuitively, longer paths require higher number of
forwarding rule lookups and thus longer lookup times. Specifically, in
Figure~\ref{fig:lookups-hopcount} the average path is 5.3 hops for
Jupiter, 5 hops for Topo-R, 4 hops for \fattree and 2.7 for
\jellyfish. 
For the same reason we observed decrease in performance (both throughput and latency) for the \texttt{Uniform plan}.



We compares \pname's lookup time with ONOS's time to compute an unprotected (single shortest path), since ONOS reactively runs a shortest path computation for each flow request. Finding a protected path is even more expensive and we leave results out. Since ONOS does not support link weights or batching lookups, we use the \texttt{Hop-count plan} with batch size of 1. 

\pname is \emph{five orders of magnitude} faster than ONOS:
$0.01ms$ for \pname vs $1850ms$ for ONOS with Dijkstra.  This should
be unsurprising: path retrieval extracting from a precomputed set 
of forwarding rules is far more time efficient than running a path
search. However, a proactive approach is only
feasible because \pname can incrementally update APSP values extremely
fast. 



%% file: 05_failures.tex
\subsection{Reacting to link  failures}\label{sec:eval_failures}
We first measure how \pname's time to react to link failures compares
with ONOS.  In particular, we explore the tradeoff between \pname's
approach of recomputing all paths incrementally in one go, and ONOS's
precomputation of single failover paths for a (protected) subset of
flows, combined with flow-by-flow recomputation for affected flows.

We first show \pname's latency in the face of a single link failure,
i.e. how fast it recomputes forwarding rules for all switches after a
link has been detected as down.  We start with a network with an
initial set of rules calculated by \pname, and then remove a
randomly-selected link from the topology.  We repeat this experiment
(with different links) 500 times, each time starting from the same
state.

We measure \pname's latency to identify and update all affected
forwarding rules.  Running with a single worker here minimizes median
latency, but we show results for 8 workers since this minimizes tail
latency.

\begin{figure*}[htb]
	\centering
	\begin{minipage}[b]{.32\textwidth}
		\includegraphics[width=.9\linewidth]{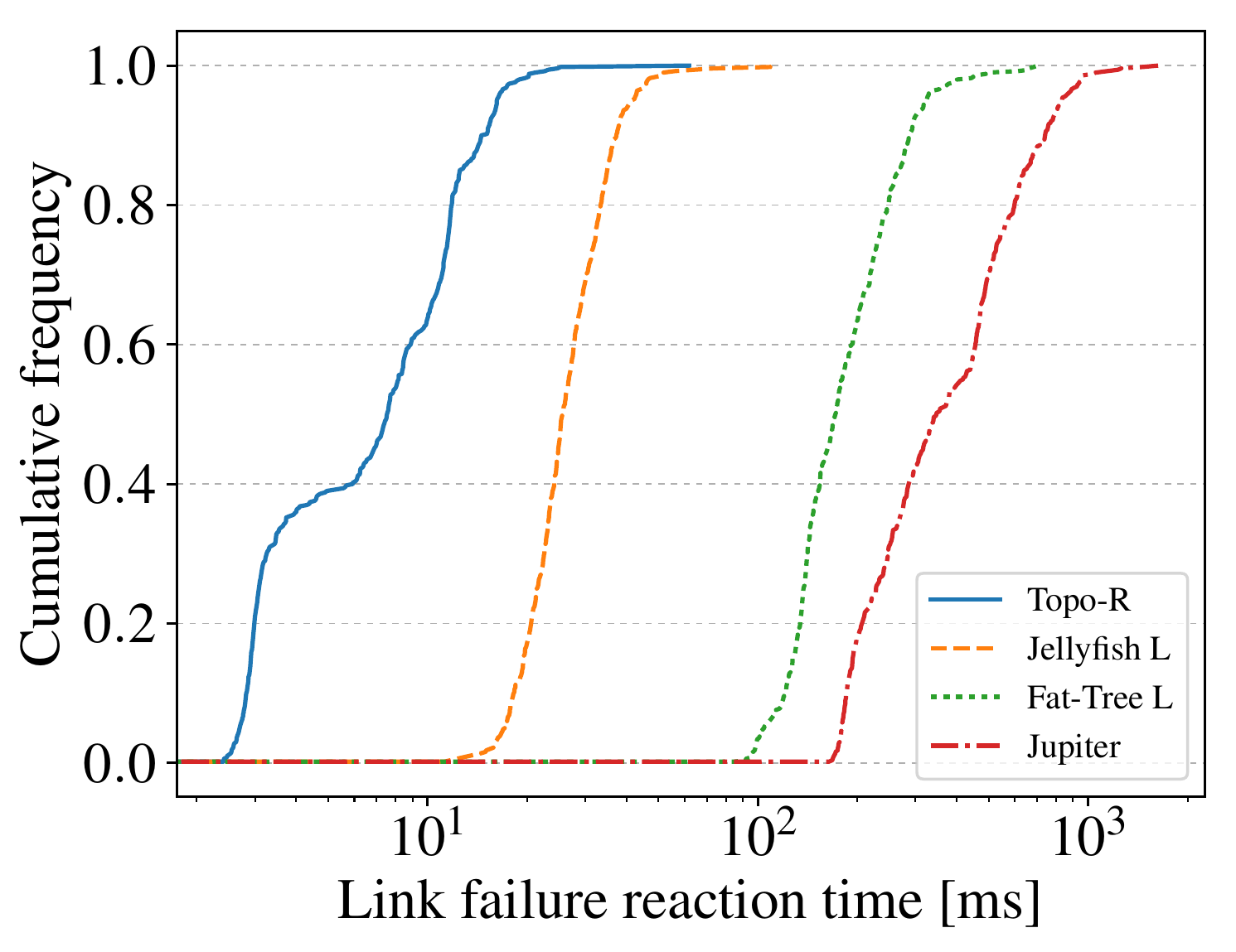}
		\caption{\small \sl  Single-link failure recovery (\texttt{Hop-count} plan).}
		\label{fig:link-failure-latency-hop-count} 
	\end{minipage}
    \hfill
	\begin{minipage}[b]{.32\textwidth}
		\includegraphics[width=.9\linewidth]{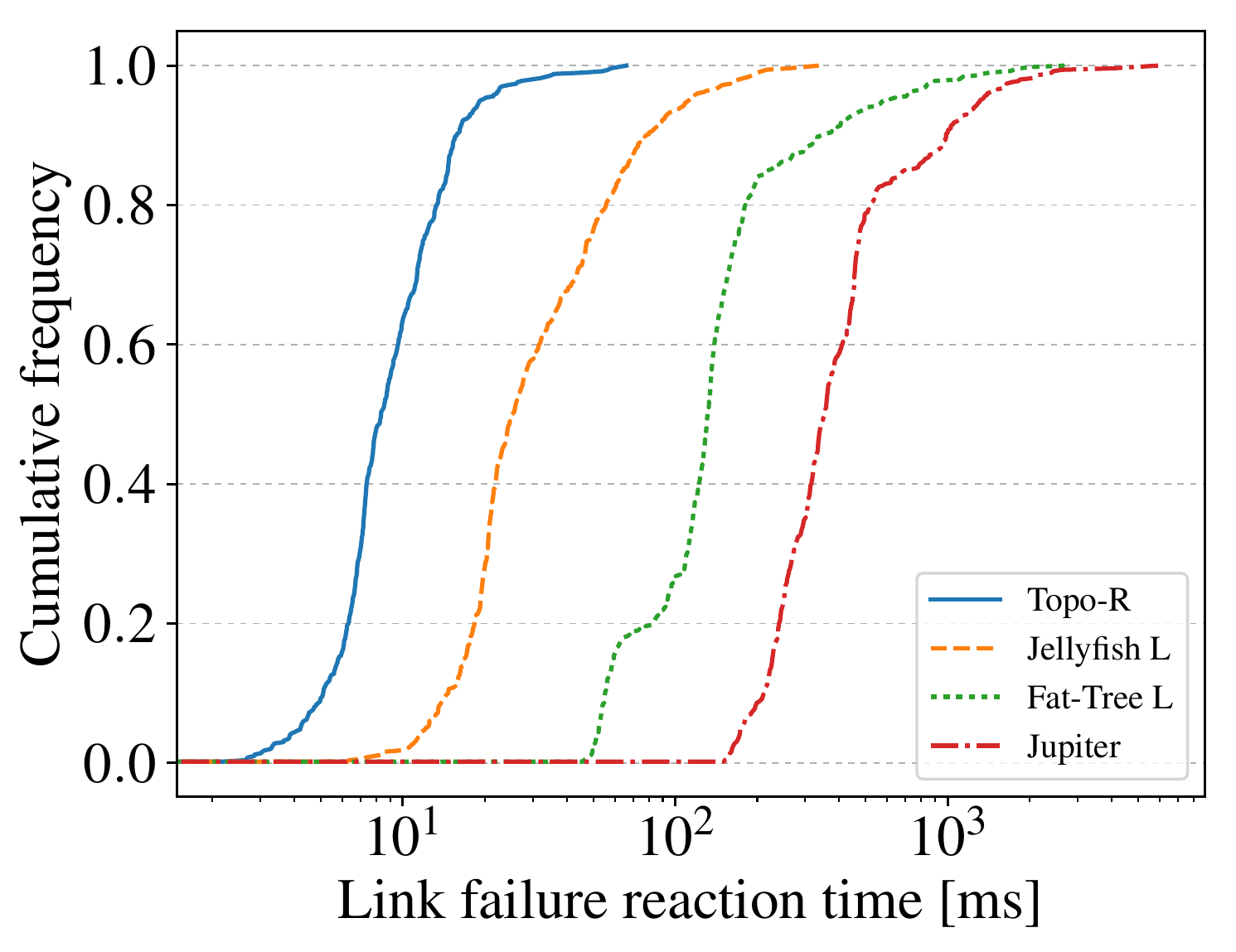}
		\caption{\small \sl  Single-link failure recovery (\texttt{Uniform} plan).}
		\label{fig:link-failure-latency-random}
	\end{minipage}\hfill
	\begin{minipage}[b]{.32\textwidth}
		\includegraphics[width=.9\linewidth]{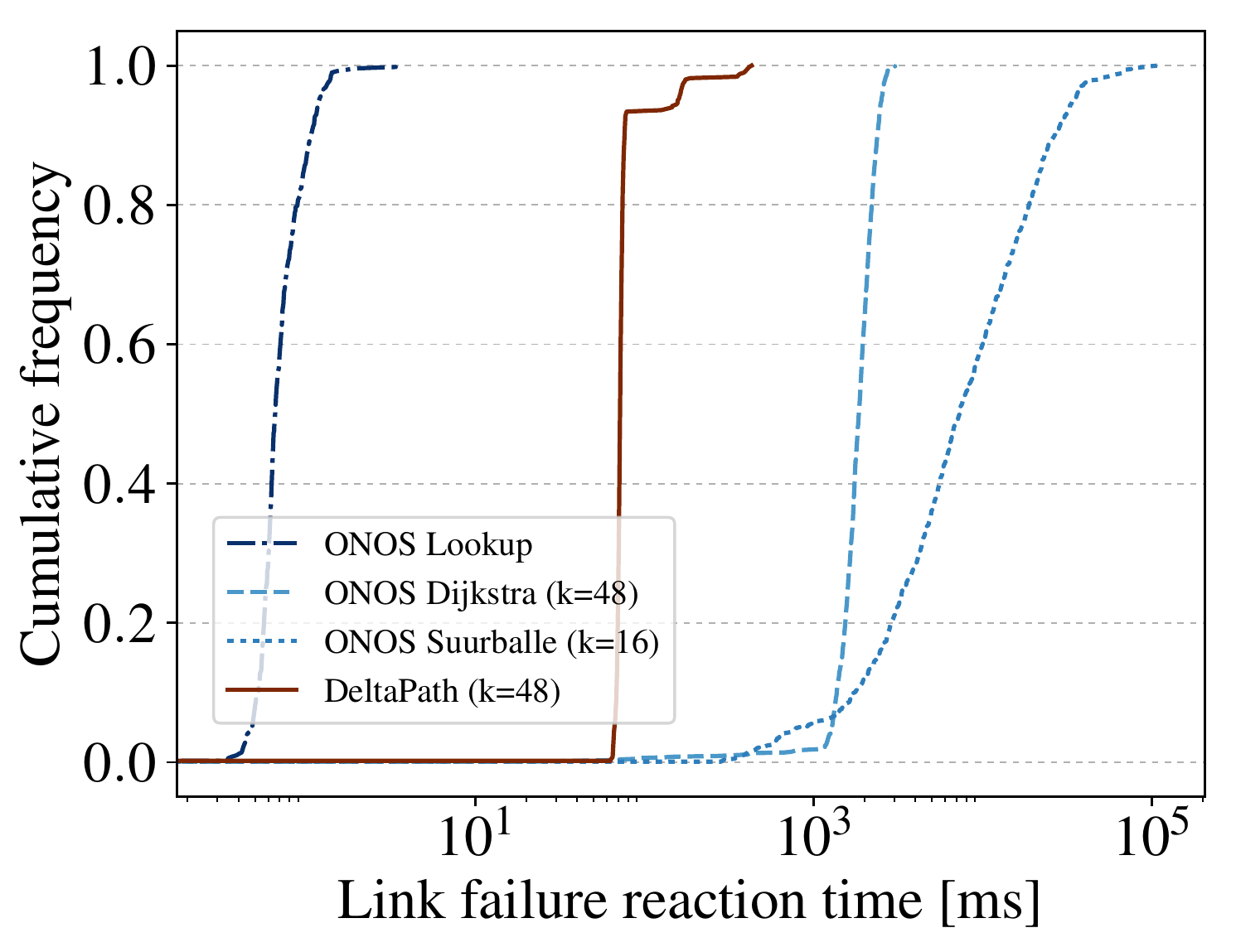}
		\caption{\small \sl  Link failure recovery for ONOS vs. \pname.}
		\label{fig:onos-failures}
	\end{minipage}\hfill	
\end{figure*}

Figures~\ref{fig:link-failure-latency-hop-count}
and~\ref{fig:link-failure-latency-random} (summarized in
Table~\ref{tab:link-failures-latencies}) show results for \pname.
Larger networks natually mean higher latencies, but median recovery
time remains below $250ms$ except for Jupiter at around $350ms$.
\pname performs better on \jellyfish than on \fattree, intuitively because \jellyfish
has smaller diameter than \fattree for the same number of hosts~\cite{Jellyfish}. 

The tails reported in Table~\ref{tab:link-failures-latencies}
result from failing links that carry many shortest paths
between pairs of switches, triggering many forwarding rule
updates.  \texttt{Uniform} shows higher latencies since its random
weights tend to introduce longer shortest paths (using sum of weights
rather than hop count). 

\begin{table}[tb]
	\setlength{\tabcolsep}{5pt} 
	\begin{center}\scriptsize
		\begin{tabular}{ c | c | c c c c }
			\hline
			& & \jellyfish & \fattree & Topo-R & Jupiter \\
			\hline
			Hop-count & Median & 25.65 & 171.94 & 7.54 & 347.28 \\ 
			plan & Worst & 114.63 & 705.87 & 61.89 & 1628.98 \\ 
			\hline
			Uniform & Median & 25.32 & 131.32 & 8.32 & 354.42 \\ 
			plan & Worst & 336.57 & 2685.66 & 66.18 & 5916.01\\ 
			\hline
		\end{tabular}
	\end{center}
	\caption{\small \sl Summary of Figures
          \ref{fig:link-failure-latency-hop-count} and  \ref{fig:link-failure-latency-random} [ms]}
	\label{tab:link-failures-latencies}
\end{table}


\textbf{Comparison with ONOS:}
Figure~\ref{fig:onos-failures} now compares \pname with
ONOS on the \fattree.  For ONOS, we generate an intent between a
random pair of switches and then fail a link that affects this intent,
repeating the experiment 500 times.  Since ONOS does not support link
weights, we compare with DeltaPath with all link weights set to 1.
Here we run \pname with a single worker, since ONOS handles each intent
on a single thread.

``DeltaPath'' shows the time for \pname to fully recompute APSP
following the link failure.  After this time, routing has effectively
been recomputed for all paths in the system.

``ONOS Dijkstra'' shows this time for ONOS to recompute the path for
the \emph{single} affected intent in the experiment, assuming it is 
unprotected and has no precomputed failover.  

This computation for just one path is, at the median, 26 times slower
than \pname and 7.6 times slower at the tail.  Multiple intents
sharing the affected link multiply this cost for ONOS, but not for
\pname.

``Protected intents'' in ONOS have precomputed failover paths and can
be restored much faster.  ``ONOS Lookup'' shows how long ONOS takes to
find each affected intent given the failed link.  After this lookup,
failing over to the precomputed backup is extremely fast -- faster than
\pname.

However, this failover latency has a cost.  If another link
affecting the backup path subsequently fails before the backup is
recomputed, ONOS must then compute SSSP for each affected intent,
which as the ``ONOS Dijkstra'' result shows can be prohibitive in cost
for links carrying many intents.

Moreover, this ``window of vulnerability'' for ONOS can be quite
large.  ``ONOS Suurballe'' shows how long ONOS takes recompute a new
backup path using Suurballe's algorithm, on a much smaller network
than the other results: ONOS runs out of memory in Suurballe on a 48-ary
fat tree, since its implementation keeps multiple copies of the entire
graph and uses extensive path materilization) so the Surballe results
here are for a 16-ary fat tree.  

With that caveat, this is the time taken for ONOS to restore path
redundancy from the point where the failure was detected, but note
again that this figure is for a single ONOS intent, and already can
run into several minutes of CPU time

The tradeoff \pname offers, therefore, is a latency of
tens-to-hundreds of milliseconds to restore complete connectivity
independent of the number of flows in the system.  This is against
ONOS' fast failover for the subset of intents protected by
(potentially prohibitively) expensive precomputation and slow
restoration of unprotected paths, both of have costs which increase
linearly in the number of flows in the system.

As an aside, \pname also compares well with distributed routing in 
Open Shortest Path First (OSPF).  An OSPF link failure causes each
switch to calculate a new routing tree with itself as root.
Convergence time after a failure is detected consists of this
computation plus a configurable delay timer (\texttt{spf-delay}),
which is set by default in Juniper switches to
\xms{200}~\cite{Juniper-SPF}, Brocade to 5s \cite{Brocade-SPF}, and
Cisco to \xms{50} (down from 5s last year)~\cite{Cisco-SPF}.

%% file: 05_switchfailures.tex
\begin{figure*}[htb]
	\centering
	\begin{minipage}[b]{.32\textwidth}
		\includegraphics[width=0.9\textwidth]{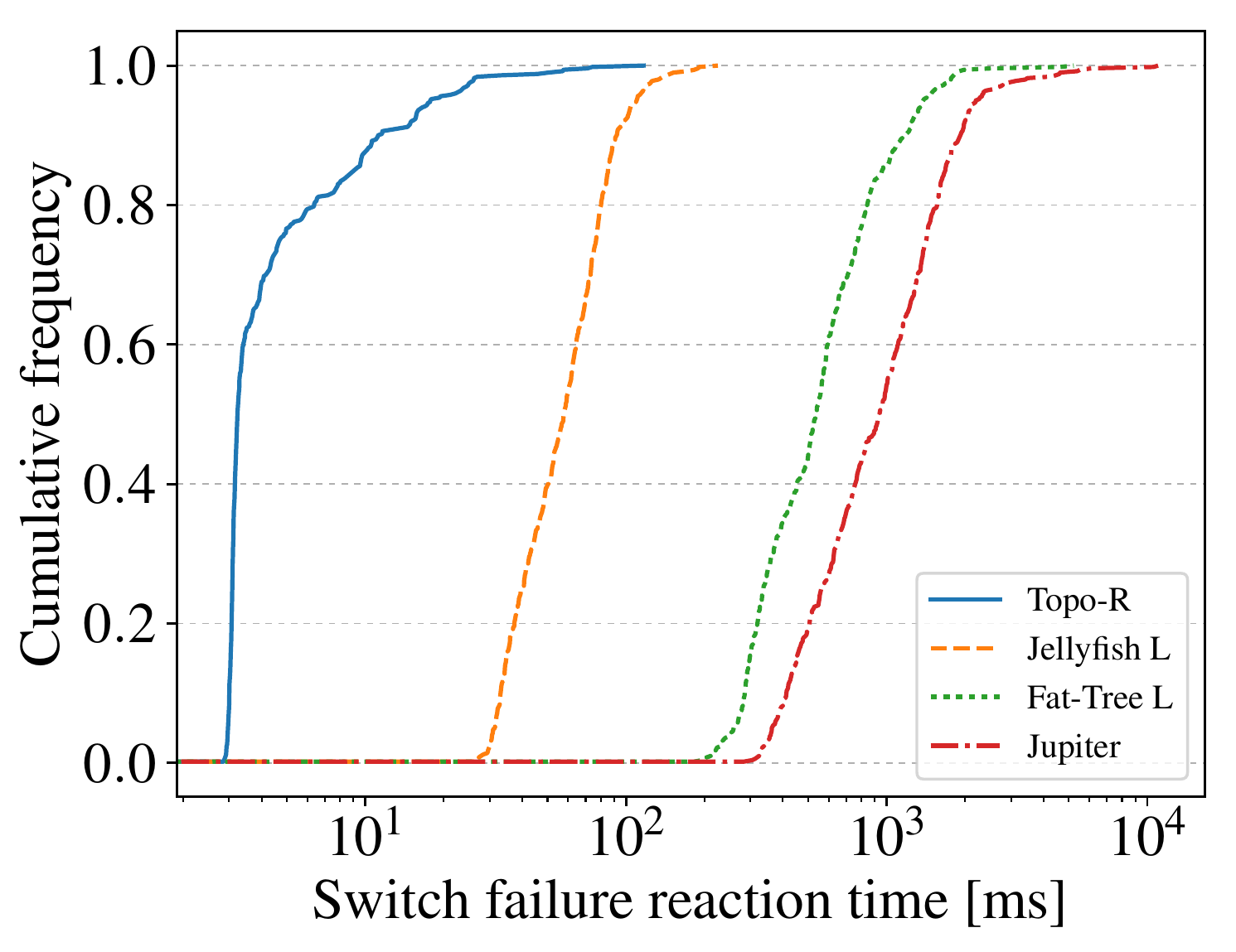}
		\caption{\small \sl Switch failure recovery (\texttt{Hop-count plan})}
		\label{fig:switch-failure-latency-hop-count}
	\end{minipage}\hfill
	\begin{minipage}[b]{.32\textwidth}
		\includegraphics[width=0.9\textwidth]{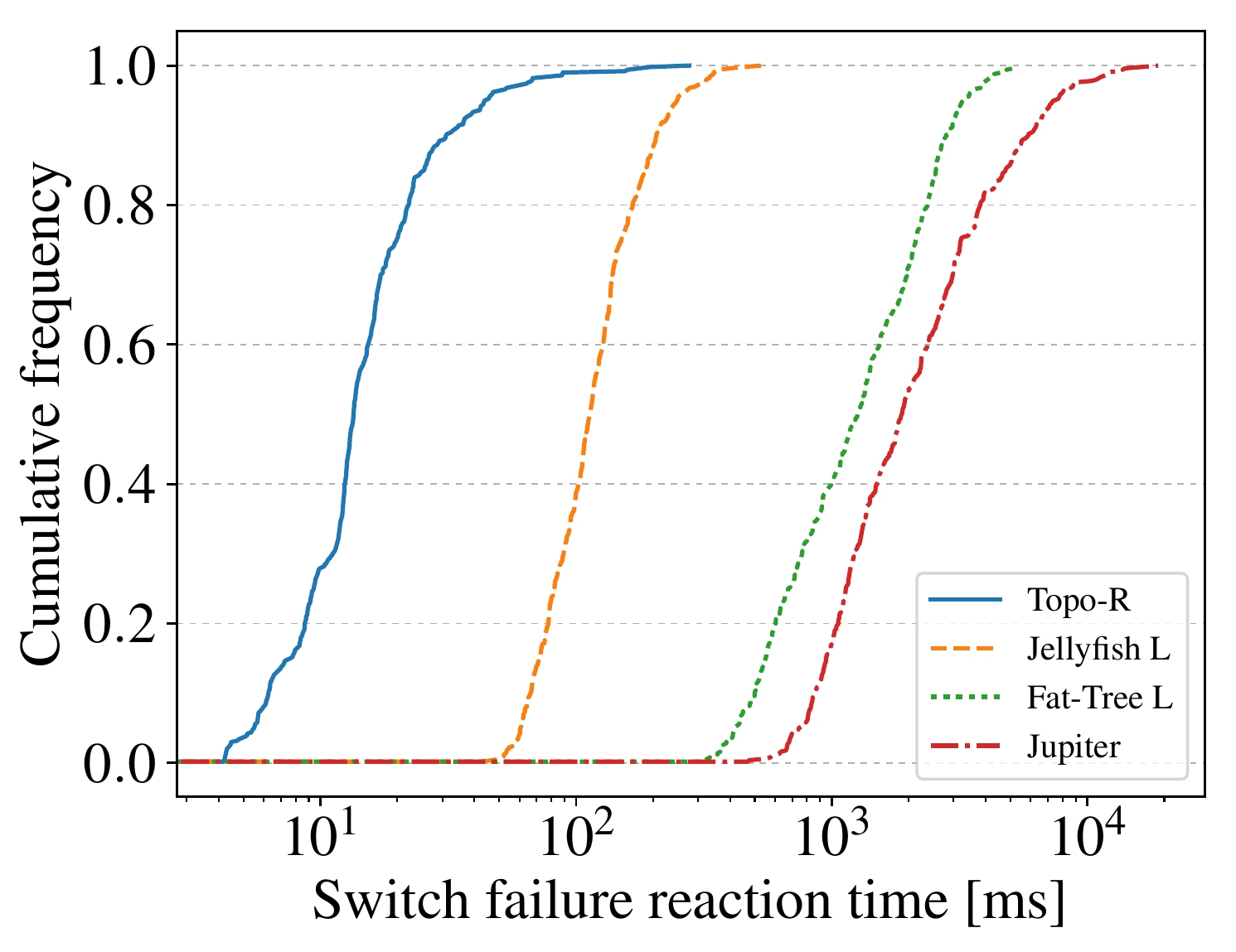}
		\caption{\small \sl  Switch failure recovery (\texttt{Uniform plan})}
		\label{fig:switch-failure-latency-random} 
	\end{minipage}\hfill
	\begin{minipage}[b]{.32\textwidth}
		\includegraphics[width=0.9\textwidth]{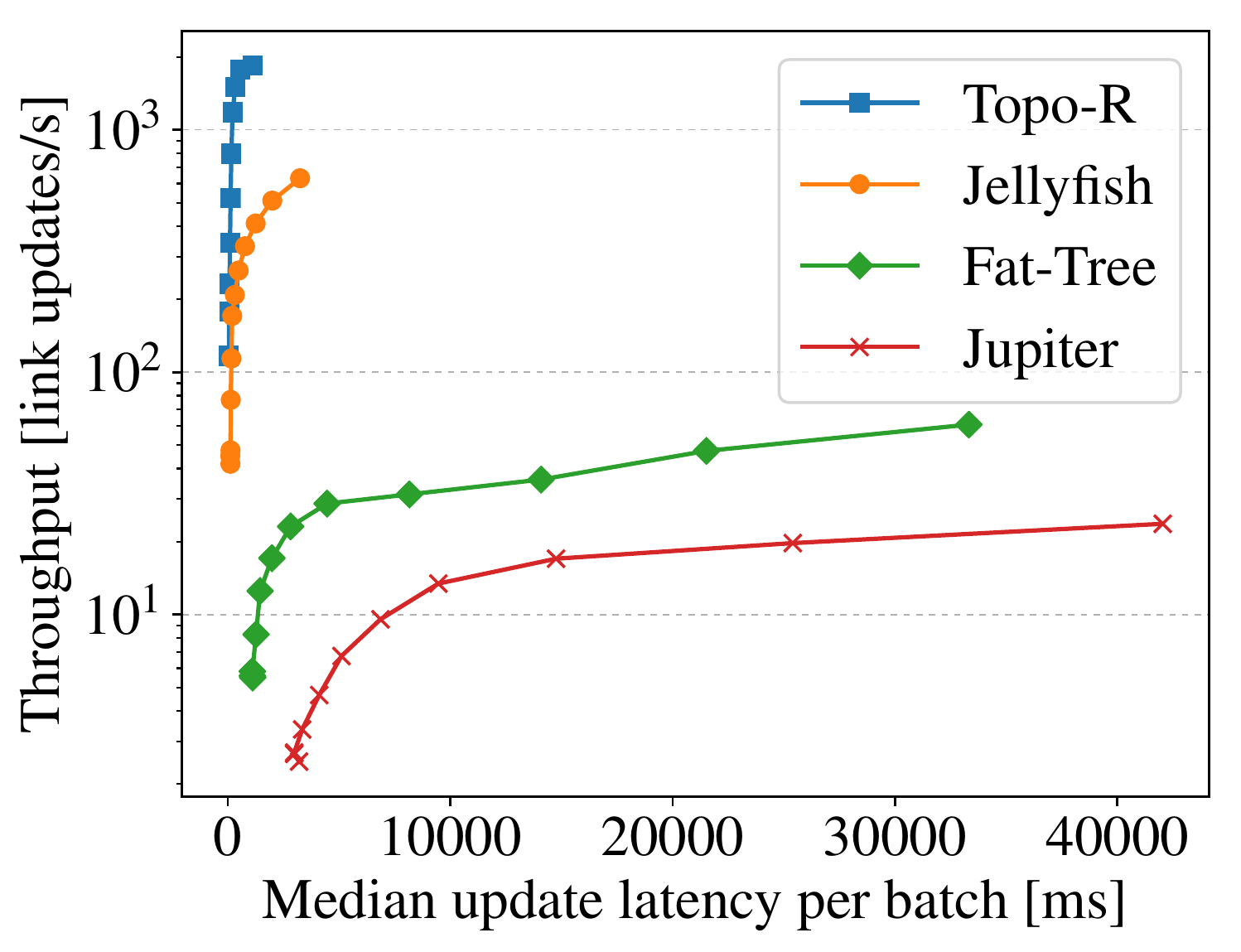}
		\caption{\small \sl Throughput vs Latency for path weight updates, (\texttt{Uniform plan})}
		\label{fig:path-update-throughput-latency}
	\end{minipage}
\end{figure*}

\textbf{Entire switch failures:} Finally, we repeat the experiment but
now fail an entire switch and its attached links. 
Figures~\ref{fig:switch-failure-latency-hop-count}
and~\ref{fig:switch-failure-latency-random}, summarized in 
Table~\ref{tab:switch-failures-latencies}, show the time take for
\pname to recover all forwarding rules. 

Performance is similar to the single-link case with a longer tail
switch failure creates more work than an average single-link failure
on the same network.  Similar arguments apply as to the relative
performance of different topologies and link weight distributions.

The comparison with ONOS is also similar.  Since ONOS pre-computes a
second, node-disjoint path for each ``protected'' intent, fast
failover will be the same for a switch failure as for a link, and so
the figures are essentially the same as in Figure~\ref{fig:onos-failures}.  
Note, however, that ONOS recomputation times are linear in the number
of intents that must be rerouted, and this figure will likely be
much higher in the case of a switch failure.

\begin{table}[htb]
	\setlength{\tabcolsep}{5pt} 
	\begin{center}\scriptsize
		\begin{tabular}{ c| c | c c c c }
			\hline
			& & \jellyfish & \fattree & Topo-R & Jupiter \\
			\hline
			Hop-count & Median & 57.69 & 534.83 & 3.22  & 939.67 \\ 
			plan & Worst & 225.13 & 5224.34 & 116.77 & 11009.44 \\ 
			\hline
			Uniform & Median & 113.13 & 1267.63 & 13.5 & 1882.75 \\ 
			plan & Worst & 537.52 & 5234.34 & 277.01 & 18921.72\\ 
			\hline
		\end{tabular}
	\end{center}
	\caption{\small \sl Summary of Figures \ref{fig:switch-failure-latency-hop-count} and \ref{fig:switch-failure-latency-random} .}
	\label{tab:switch-failures-latencies}
\end{table}

\textbf{Discussion:} 
\pname's performance in this experiment is due both to an 
incremental algorithm and a high-performance implementation. 
Govindan et al.~\cite{Govindan16} report that
80\% of failures in \jupiter take between 10 and 100 minutes. This
includes the manual task of restoring physical connectivity but it may
be still interesting to map \pname's performance to the lower
bound. From Table~\ref{tab:link-failures-latencies}, median recovery
time for Jupiter topology with hop-count (equal link weights)
constitutes 3\% from the reported 10min lower bound, while worse-case
performance constitutes 16\%.

%% file: 05_updates.tex
%

\subsection{Link weight updates}\label{sec:eval_updates}

The merits of \pname's incremental computation apply also to handling concurrent link weight updates. 
Reacting quickly to large numbers of link weight updates is important for flexible network control, one example being bandwidth-constraint routing which we address in \cref{sec:qos-eval}. 
Here, we focus on the throughput-vs-latency trade-off \pname achieves in handling weight updates, using the \texttt{Uniform plan}.

In this experiment, we form batches of link weight updates which mimic addition and removal of host-to-host flows and their resource consumption. We (i) randomly select a path and add its links to the batch, and (ii) link weight updates are fixed at 5\% of total link capacity, representing the maximum flow size seen in public traces \cite{fb_trace,MS_trace}. Taking a conservative number let us establish an upper bound on performance rather than tailor results to individual workload distributions. We apply 500 such batches successively to a network initialized as in \cref{sec:eval_failures}, making updates in this experiment
\emph{additive}. \pname uses 32 worker threads. 



\begin{table}
	\setlength{\tabcolsep}{5pt} 
	\begin{center}\scriptsize
		\begin{tabular}{ c|c|c c c c }
			\hline
			 Metric & Target (s) & \jellyfish & \fattree & Topo-R & Jupiter \\
			\hline
			median throughput & 1 & 350 & 5 & 1800 & - \\ 
			& 10 & 973* & 33 & 3230* & 13 \\
			\hline
			mean path length & & 5 & 6.4 & 5 & 8 \\
			\hline
			\multicolumn{6}{l}{(*) extrapolated result due to insufficient topology size to extract large pool of paths. }
		\end{tabular}
	\end{center}
	\caption{\small \sl Throughput of processed link weight updates for a target $1s$ latency.}
	\label{tab:updates}
\end{table}

Figure~\ref{fig:path-update-throughput-latency} shows throughput
vs. median latency for applying updates as the batch size varies
Each marker represents a batch size between 1 and 1028 in powers of two. 
Table~\ref{tab:updates} further extracts the median throughput and corresponding
average path length (in hop count) for both plans with a target latency
of \xs{1}.   

\pname achieves the \xs{1} latency target at high throughput on all
topologies except Jupiter, where it required median \xs{3}
for an 8-hop path. At a \xs{10} latency target, throughput grows to 13 updates/s in Jupiter and 20 updates/s in \fattree. 

Convergence time is the main factor in \pname's performance: large well-connected graphs like Jupiter and \fattree need higher number of iterations to converge due to longer average path length (cf. Table~\ref{tab:updates}). With twice as many nodes, Jupiter sees bigger number of affected paths. 

To our knowledge, \pname is the first system to handle concurrent network changes efficiently. Most open source SDN routing modules restart the computation, which does not scale. Despite its popularity, ONOS does not support arbitrary link weight updates. If we were to enable it, in the time ONOS would update a single intent ($1850ms$), \pname can
update \emph{all} rules affected by 32 link weight updates. 




%% file: 05_policy.tex
\subsection{QoS routing}\label{sec:qos-eval}

We now show how \pname behaves for all routing algorithms of Table~\ref{tbl:functions}. 
For each algorithm, we first initialize \pname with \texttt{Uniform} plan (except in the case of hop-based where we use \texttt{Hop-count} plan), and let it compute an initial set of base rules.
Then, we generate 1000 flow requests, which are submitted one after the other to \pname, and we measure \pname's latency to retrieve the best path and update its collection of forwarding rules. 
Flows' origin and target nodes are randomly chosen, whereas their sizes and inter-arrival times are based on the distributions extracted from the publicly available Facebook traces in \cite{fb_trace} (for cache leaders). The collection of base rules is updated after each request by all algorithms except the hop-count. 
This is done to keep the graph snapshot consistent for the next request, and reflects the fact that link weights change according to the active flows and their size (i.e. the bandwidth they reserve).

\begin{figure}
	\centering
	\includegraphics[width=.9\linewidth]{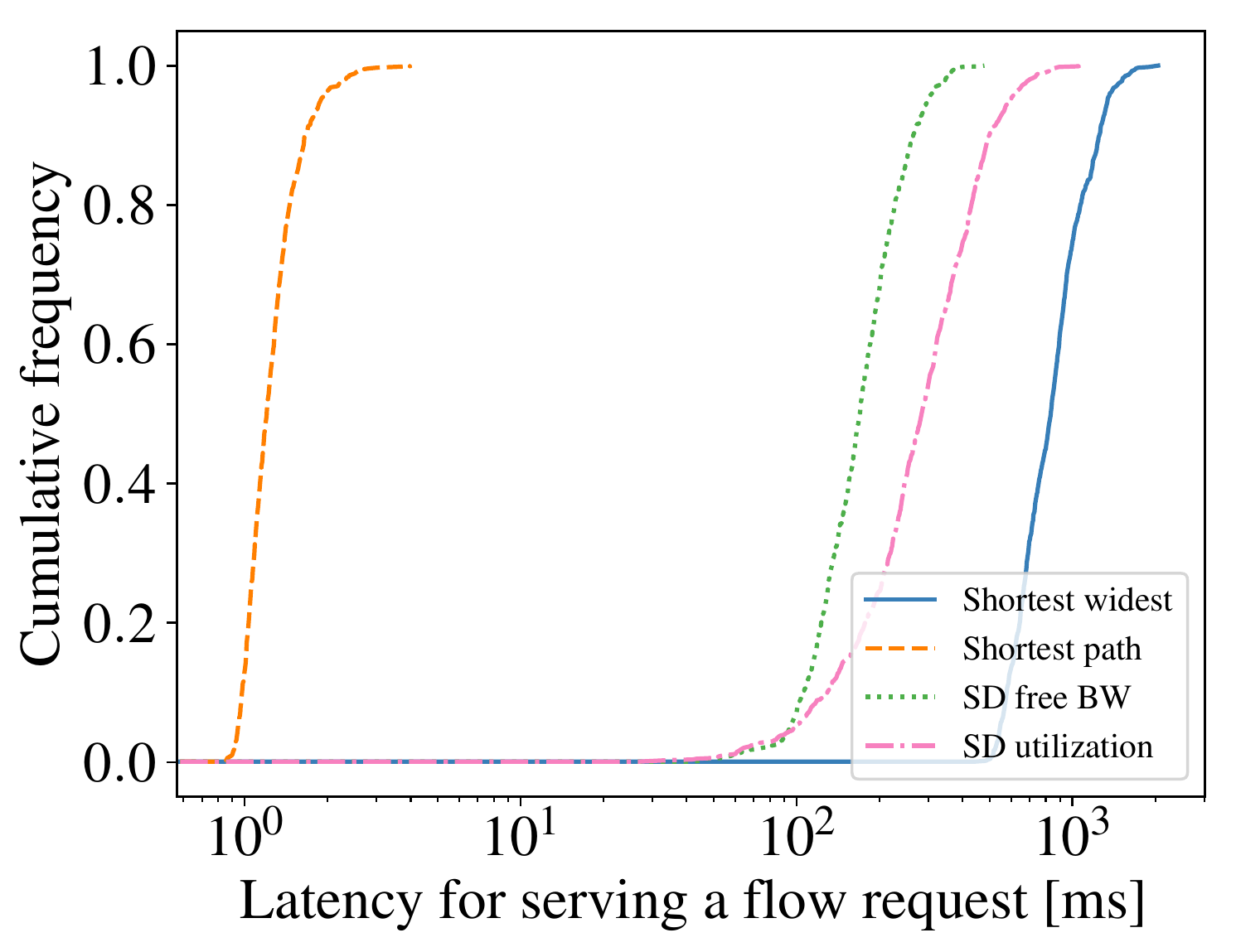}
	\caption{\small \sl Latency for serving flow requests (including path retrieval and rule updates) on Jupiter. }\label{fig:bw}
\end{figure}

Figure~\ref{fig:bw} shows \pname's latency distribution for each algorithm on Jupiter (results are similar for other topologies). Shortest path (hop-based) is the fastest since it is workload insensitive, while shortest-widest path algorithm has higher latencies because the paths it updates tend to be longer than in other algorithms. The two shortest distance algorithms react faster with SD free BW (relating weights to available link bandwidth) being the faster one. This is explained with the difference in link cost functions: the linear cost function of SD utilization (relating weights to link utilization) causes even small flows as those in \cite{fb_trace} to trigger rules updates, while this is not the case for SD free BW.

We conclude that \pname can offer performance benefits to different algorithms which can be expressed in its model. 






\subsection{Policy evaluation}\label{sec:policy_eval}

Here we evaluate \pname's performance on policy evaluation, as described in \cref{sec:policy_implementation}.
For the experiments of this section we only use Jupiter, which is the largest of all topologies in Table~\ref{tab:topology-params}. 
Performance results with other topologies are similar and are omitted due to lack of space.

\begin{figure}
	\centering
	\includegraphics[width=.9\linewidth]{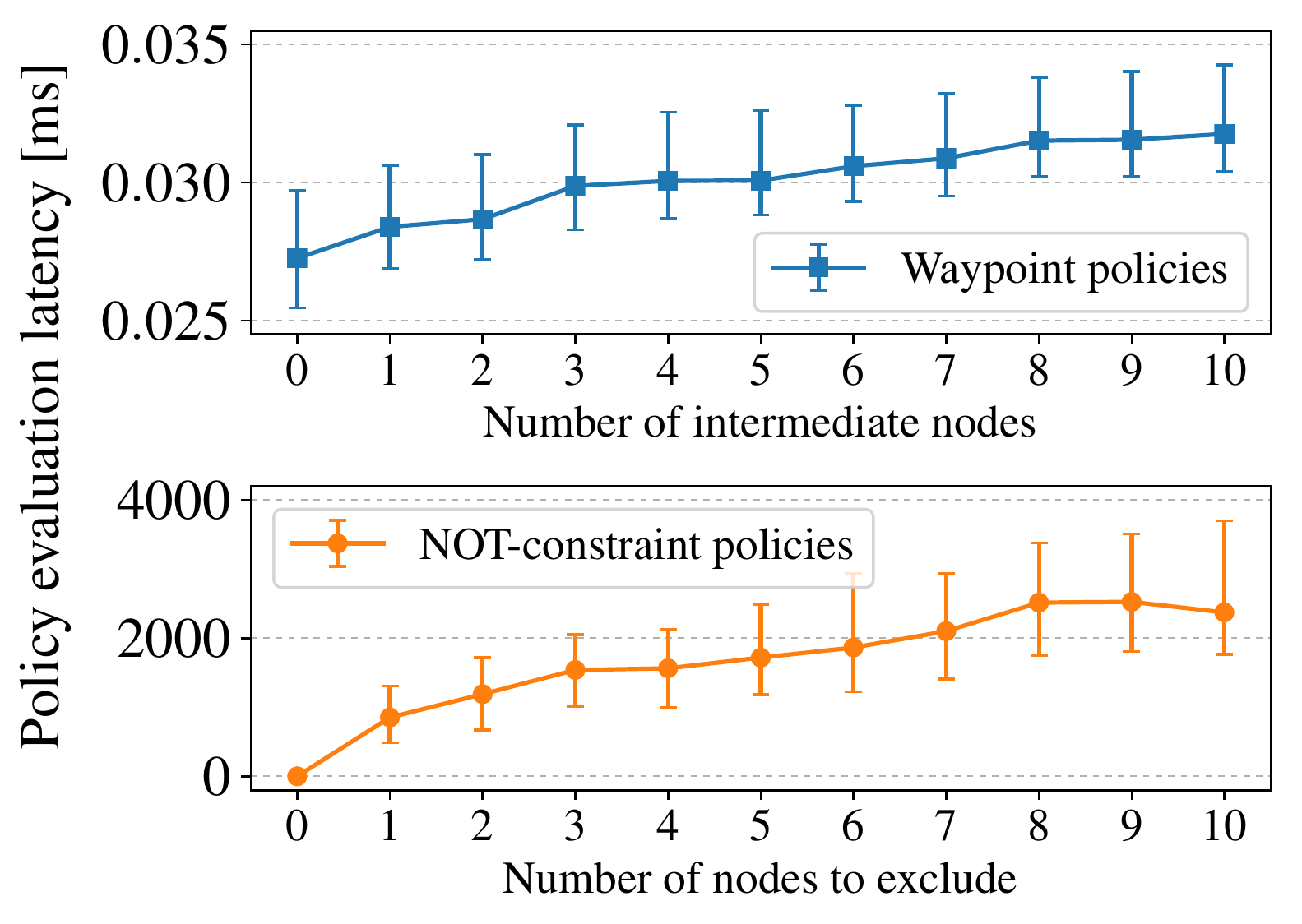}
	\caption{\small \sl Latency for policy evaluation on Jupiter. Policies without constraints ($k=0$) correspond to a single path retrieval and are included here for reference. }\label{fig:policy-evaluation}
\end{figure}

\stitle{Waypointing policies.} Figure~\ref{fig:policy-evaluation} (top) shows \pname's latency in evaluating waypointing policies on Jupiter. 
In these experiments we vary the number of intermediate nodes the path has to go through from $k=0$ to $k=10$ ($x-axis$). 
For each value of $k$, we generate 500 policies by randomly choosing the origin, target and intermediate nodes among all nodes in the topology. 
Each box-plot shows \pname's latency distribution in evaluating 500 random policies independently.
Since waypoiting policies are reduced into one or more fast path retrievals (cf. \cref{sec:path_retrieval}), latency in Figure~\ref{fig:policy-evaluation} (top) remains low and increases slightly with the number of intermediate nodes. 

\stitle{NOT-constraint policies.}
Figure~\ref{fig:policy-evaluation} (bottom) shows \pname's latency in evaluating NOT-constraint policies. 
In these experiments we vary the number of intermediate nodes to exclude from a path from $k=0$ to $k=10$ ($x-axis$). 
For each value of $k$, we generate 500 NOT-constraint policies by randomly choosing the origin, target and intermediate nodes as before. 
Each box-plot shows \pname's latency distribution in evaluating 500 random policies independently.
All node removals per policy are submitted in a single batch to the routing component and are processed in parallel. 

The time to evaluate NOT-constraint policies is significantly larger compared to waypointing policies because, in this case, the policy component needs to wait for the routing component
to compute the new base rules after the node removals  (cf. \cref{sec:policy_implementation}). 
Note that the results in Figure~\ref{fig:policy-evaluation} (bottom) are consistent with those in Figures \ref{fig:switch-failure-latency-hop-count} and \ref{fig:switch-failure-latency-random} for Jupiter.


%% file: 05_bench_parallelism.tex
\subsection{Scalability}\label{sec:microbenchmarks}

We finally examine how \pname's throughput and latency scale with 
worker thread count. For most cases, \pname scales respectably, and
can show 6-fold speedup on Jupiter with 32 worker threads.

We initialized each topology with the \textit{Uniform} plan and
measured throughput and latency over 1000 runs while increasing the
update batch size from 1 to 512 in powers of two, and the number of
worker threads from 1 to 32 (the maximum number of hardware contexts
on the machine). 

\begin{table}[htb]
	\setlength{\tabcolsep}{5pt}
	\begin{center}\scriptsize
		\begin{tabular}{ c | c c c c c c }
			\hline
			\#workers & 1 & 2 & 4 & 8 & 16 & 32 \\
			\hline
			\multirow{2}{*}{Fat-Tree} & 26 & 24 & 24 & 30 & 56 & 77\\ 
			& (37.91) & (41.09) & (40.87) & (32.50) & (17.62) & (12.97)\\ 
			\hline
			\multirow{2}{*}{Jellyfish} & 174 & 212 & 198 & 188 & 337 & 426\\ 
			& (5.50) & (4.70) & (5.03) & (5.29) & (2.96) & (2.34)\\
			\hline
			\multirow{2}{*}{Topo-R} & 492 & 566 & 656 & 1142 & 1343 & 1445 \\
			& (2.02) & (1.76) & (1.52) & (0.87) & (0.74) & (0.69)\\ 
			\hline
			\multirow{2}{*}{Jupiter} & 8 & 10 & 14 & 19 & 34 & 53  \\
			& (123.14) & (96.00) & (70.28) & (51.08) & (29.32) & (18.80) \\ 
			\hline
		\end{tabular}
	\end{center}
	\caption{\small \sl Maximum throughput vs. parallelism, \texttt{Uniform}
          plan.  Optimal batch sizes are shown, with latency (msec) per
          link weight update below in parentheses.}
	\label{tab:scaling}
\end{table}




Table~\ref{tab:scaling} shows maximum median throughput
and latency (in parentheses) for varying degrees of
parallelism, for the optimal batch size in each case. 


Such measurements should be treated with caution, but for many
topologies we can conclude that \pname scales well once the number of
updates is large enough to make the computation CPU bound.  This fits
intuitively with the data-parallel computation of the routing operator.

%% file: 07_deployment.tex
\section{Deployment considerations}\label{sec:deployment}

\pname is designed as a routing application easy to deploy on any SDN controller. The one requirement we impose is the streams format and semantics (network updates, policy updates, flow rules) to be consistent with the controller API.

\stitle{Routing granularity.} \pname's execution model is agnostic to routing granularity. The technique applies equally well to computing per-flow forwarding rules or finding paths between ToR switches. Host-based and switch-based routing are handled by the same \pname computation, where the latter simply requires mapping the produced rules, containing switch addresses, to IP addresses.
\pname allows network architects to make routing granularity decisions.

\stitle{Dynamic QoS routing.} In real networks, several \pname instances may be deployed in parallel, each running a distinct QoS routing strategy (\cref{sec:qos_implementation}). The behavior of those instances depends on whether the routing strategy uses static or dynamic weights. With static weights, forwarding rules can be pre-installed at switches and controller intervenes only upon changes in network topology or administrator policies. With dynamic weights however, routing decisions adapt continuously to live performance metrics, e.g., link utilization, and pre-installing rules is not feasible. Flows receive performance guarantees at the cost of inflated setup times (caused by both controller processing and frequent rule updates in switches). 

In practice, only a subset of network flows would require dynamic routing. Interactive applications in modern-day datacenters (e.g., web and OLTP) have stringent performance requirements in terms of delivered throughput and corresponding low tail latency \cite{DeJ13}. Parley \cite{JeV14} and \cite{JeL14} for Wikipedia benchmark \cite{wiki} demonstrate that bandwidth provisioning has strong impact on tail latency. Based on results in \cref{sec:eval_updates}, we believe \pname could offer an alternative scalable runtime mechanism to enforce bandwidth guarantees for interactive applications. We leave those investigations for future work. 

\stitle{Multipath routing.} \pname currently discovers all paths between pairs of nodes and uses the path selection function $f_s$ to only keep one. \pname could support ECMP-like load balancing by simply adapting the implementation to keep all discovered paths. An alternative research direction which we currently pursuit extends \pname's execution model towards algorithms for disjoint shortest path~\cite{Suurballe}.  


%% file: 02_relatedwork.tex
\section{Related work}\label{sec:related_work}

\stitle{Traffic engineering.}
Traffic engineering (TE) is an active and busy field of research. One approach, widely used in practice for its simplicity, is to orchestrate link weights of distributed protocols such as OSPF and its multipath enhancement ECMP \cite{Chiesa16,Fortz00,Fortz02}. A drawback is long convergence time after failures or updates to link weights. Centralized network control can only partially alleviate the latency problem \cite{Liu14}, for which reason recent TE proposals compute sets of paths offline \cite{Liu16,Suchara11}, infrequently upon demand~\cite{Hong13} or topology change~\cite{Kumar16}, or in a hierarchical process starting at hosts \cite{Kumar15}.
Our approach does not target TE design but complements it by enabling faster path selection, getting us closer to dynamic centralized TE. 

%


\stitle{Shortest path algorithms.}
In traditional routing, \gls{OSPF} has long supported an incremental
optimization~\cite{McQuillan80} (as has IS-IS~\cite{ISIS}).  
In contrast, the centralized computation and single network view in
SDNs favors all-pair shortest path algorithms (\gls{APSP}), where an incremental approach is harder.
King~\cite{King99} proposed the first incremental \gls{APSP} algorithm
to outperform from-scratch recomputation. Later, \cite{Demetrescu01,Demetrescu03} show practicality for large networks.

Algorithmic optimizations from the database and data mining communities as those in \cite{Akiba14,DAngelo16,Raghavendra12} and \cite{Hayashi16} can be additionally applied to our approach. Bi-directional heuristic search \cite{Kaindl97} deserves exploration too. 

\stitle{SDN routing modules.}
Current SDN controller \emph{implementations} lag
behind the algorithms. Most SDN routing applications are based on pure single-source shortest path (\gls{SSSP}) routing logic. 
SDN routing applications derived from academic projects~\cite{POX, Ryu} typically implement basic, unoptimized algorithms~\cite{Berde14}.
ONOS~\cite{ONOS} and OpenDaylight~\cite{Opendaylight} are more
modular and flexible (e.g.\ \gls{ONOS} supports both Dijkstra and
Suurballe's disjoint shortest paths). In contrast, OpenMUL \cite{OpenMUL}, another production-ready controller, implements \gls{APSP}. All open source SDN routing modules we are aware of rely on non-incremental algorithms. We however show that an incremental approach has significant performance advantages. 

\stitle{Incremental graph processing.}
Incremental graph processing approaches already exist:
Kineograph~\cite{Kineograph} ingests graph updates in batches to
construct a series of consistent graph
snapshots. Chronos~\cite{Chronos} optimizes this using locality-aware
scheduling across multiple graph snapshots. However, these
batch-oriented systems are designed for high throughput but not low
latency, a requirement for SDN route computation.  


%% file: 08_conclusion.tex
\section{Conclusion}\label{sec:conclusion}

Representing routing computations as an online, incremental
fixed-point computation on streams of network and policy updates is a
promising option for scalable SDN controllers.

Combining this model with a high-performance implementation based on
\timely, our prototype SDN routing application for all-pairs shortest path,
\pname, delivers much higher performance than existing controllers,
and shows that is feasible to maintain APSP data even under highly
dynamic conditions such as frequent changes in link attributes.

\pname's performance shows that an execution model which recasts route
calculations as incremental dataflow computation can change the design
space for centralized datacenter network control.